\documentclass[aps, prb, superscriptaddress, reprint, twocolumn, a4, notitlepage]{revtex4-1}
\usepackage{graphicx}
\usepackage{amsmath,bm}
\usepackage{dcolumn}
\usepackage{mathrsfs}
\usepackage{physics}
\usepackage[utf8]{inputenc}

\usepackage{color}

\newcommand{\mr}{\mathbf{r}}

\newcommand{\mE}{\mathbf{E}}
\newcommand{\mEh}{\mathbf{\hat{E}}}

\newcommand{\mG}{\mathbf{G}}
\newcommand{\rev}[1]{#1}
\newcommand{\revv}[1]{#1}
\newcommand{\pk}[1]{#1}

\newcommand{\mft}{\tilde{\mathbf{f}}}
\newcommand{\mFt}{\tilde{\mathbf{F}}}
\newcommand{\tlo}{\tilde{\omega}}
\newcommand{\ptk}[1]{#1}

\begin{document}
\title{Quantum theory of {two-dimensional} materials coupled to {electromagnetic resonators}}\date{\today}

\author{Emil V. Denning}
\email{emil.denning@gmail.com}
\affiliation{Department of Photonics Engineering, Technical University of Denmark, 2800 Kgs. Lyngby, Denmark}
\affiliation{NanoPhoton - Center for Nanophotonics, Technical University of Denmark, Ørsteds Plads 345A, DK-2800 Kgs. Lyngby, Denmark}
\affiliation{Nichtlineare Optik und Quantenelektronik, Institut f\"ur Theoretische Physik, Technische Universit\"at Berlin, 10623 Berlin, Germany}

\author{Martijn Wubs}
\affiliation{Department of Photonics Engineering, Technical University of Denmark, 2800 Kgs. Lyngby, Denmark}
\affiliation{NanoPhoton - Center for Nanophotonics, Technical University of Denmark, Ørsteds Plads 345A, DK-2800 Kgs. Lyngby, Denmark}
\affiliation{Centre for Nanostructured Graphene, Technical University of Denmark, 2800 Kgs. Lyngby, Denmark}

\author{Nicolas Stenger}
\affiliation{Department of Photonics Engineering, Technical University of Denmark, 2800 Kgs. Lyngby, Denmark}
\affiliation{NanoPhoton - Center for Nanophotonics, Technical University of Denmark, Ørsteds Plads 345A, DK-2800 Kgs. Lyngby, Denmark}
\affiliation{Centre for Nanostructured Graphene, Technical University of Denmark, 2800 Kgs. Lyngby, Denmark}

\author{Jesper M\o rk}
\affiliation{Department of Photonics Engineering, Technical University of Denmark, 2800 Kgs. Lyngby, Denmark}
\affiliation{NanoPhoton - Center for Nanophotonics, Technical University of Denmark, Ørsteds Plads 345A, DK-2800 Kgs. Lyngby, Denmark}

\author{Philip Trøst Kristensen}
\affiliation{Department of Photonics Engineering, Technical University of Denmark, 2800 Kgs. Lyngby, Denmark}
\affiliation{NanoPhoton - Center for Nanophotonics, Technical University of Denmark, Ørsteds Plads 345A, DK-2800 Kgs. Lyngby, Denmark}

\date{\today}

\begin{abstract}
{We present a microscopic quantum theory of light-matter interaction in pristine sheets of two-dimensional semiconductors coupled to localized electromagnetic resonators such as optical nanocavities or plasmonic particles. The light-matter interaction breaks the translation symmetry of excitons in the two-dimensional lattice, and we find that this symmetry-breaking interaction leads to the formation of a localized exciton state, which mimics the spatial distribution of \pk{the} 
electromagnetic field of the resonator. \pk{The} 
localized exciton  \pk{state} is in turn coupled to \pk{an} environment of residual exciton \pk{states}. We quantify the influence of \pk{the environment} 
and find that \pk{it is} 
most pronounced for small lateral confinement length scales of the \pk{electromagnetic field in the} resonator, and \pk{that environmental effects} can be neglected if this length scale is sufficiently large. \pk{The} 
microscopic theory provides a physically appealing derivation of the coupled-oscillator models widely used to model experiments on these types of systems, \pk{in which all} 
observable quantities are directly derived from the material parameters and the properties of the \pk{resonant electromagnetic field.
As a consistency check, we show that the theory recovers the results of 
semiclassical electromagnetic calculations and experimental measurements of the excitonic dielectric response in the linear excitation limit. The theory, however, is not limited to linear response, and in general 
%
describes nonlinear exciton-exciton interactions in the localized exciton state, thereby providing a powerful means of investigating the 
nonlinear optical response of such systems.}
}

\end{abstract}
\maketitle


\section{Introduction}
\label{sec:introduction}

Over the last decade, there has been a growing interest in excitonic properties of two-dimensional (2D) semiconductors, especially in monolayers of the transition-metal dichalcogenide family~\cite{wang2018colloquium}. Owing to a direct bandgap in the visible frequency range and large exciton binding energies, these materials are particularly interesting for polaritonic physics and technology~\cite{sanvitto2016road}. Indeed, pristine sheets of these materials have been interfaced with optical nanocavities or plasmonic resonators, leading to coupling strengths of the order of 100 meV~\cite{wen2017room,zheng2017manipulating,kleemann2017strong, cuadra2018observation, stuhrenberg2018strong, han2018rabi, geisler2019single, qin2020revealing}. Different models have been used to describe the experiments and to account for the fact that the high interaction strengths have been reached even without the need for careful positioning of the nano-resonator to align it with a local defect in the 2D material. Phenomenological treatments originating in the quantum optics literature 
view the \pk{electronic excitations} 
in the 2D-material that couple to the \pk{resonant electromagnetic field} as a collection of $N$ independent dipole particles, where $N$ is typically fitted to match the experimentally observed light-matter coupling strength~\cite{han2018rabi, stuhrenberg2018strong, wen2017room, cuadra2018observation, qin2020revealing}. Within such $N$-dipole theories, the effective light--matter coupling strength \pk{is proportional to  
$\sqrt{N/V}$}, where $V$ is \pk{an effective electromagnetic mode volume}. \pk{Although the phenomenological models can be well fitted to experimental data, there appears to be no convincing microscopic theory explaining the origin or nature of the dipolar particles in the seemingly pristine 2D materials. As a consequence, the usefulness of the quantum optical concepts has been questioned~\cite{tserkezis2020applicability}, and it has been argued that a framework rooted in the condensed-matter theory of quantum wells is more appropriate. In particular, it has been recognised that such an approach can lead to models in which the electromagnetic resonator couples to a single effective exciton state~\cite{verger2006polariton, tserkezis2020applicability}.}


\begin{figure}
  \centering
  \includegraphics[width=\columnwidth]{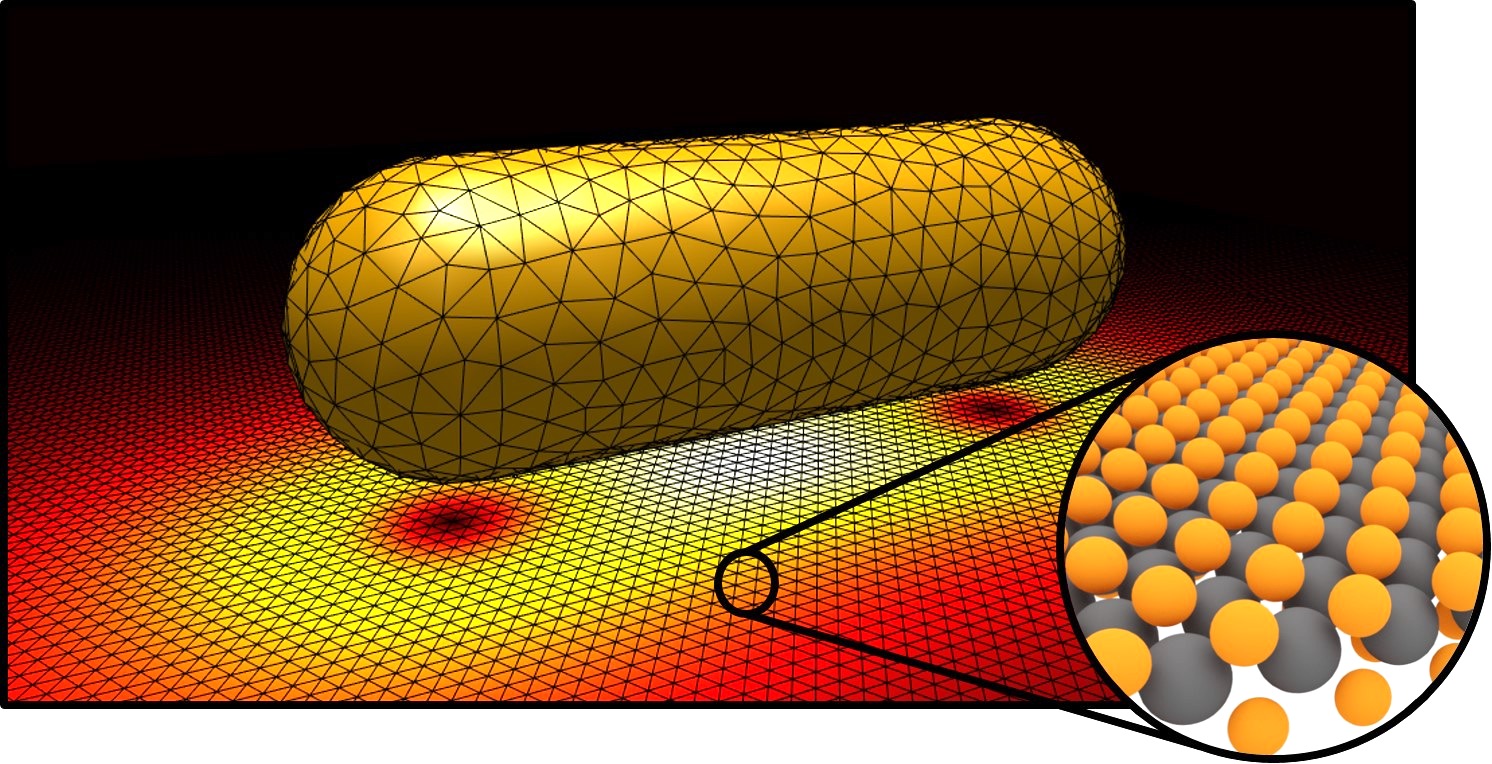}
  \caption{Illustration of an electromagnetic resonator in the form of a plasmonic nanorod situated above an infinite sheet of 2D semiconductor material. Light-matter interaction between the resonant electromagnetic field and excitons in the semiconductor leads to the formation of a localized collective exciton mode with a center-of-mass wave function matching the in-plane electric field profile indicated by the color coding.}
  \label{fig:intro}
\end{figure}

In this \pk{article} 
we build on similar ideas as put forward in Ref.~\onlinecite{tserkezis2020applicability} and develop a microscopic quantum theory for excitons in 2D materials coupled to electromagnetic resonators (see Fig.~\ref{fig:intro}). The resonator may be realized using plasmonic resonances in metal~\cite{wen2017room,zheng2017manipulating,kleemann2017strong,cuadra2018observation,stuhrenberg2018strong,han2018rabi,geisler2019single,qin2020revealing} or dielectric nanocavities~\cite{wu2014control,noori2016photonic,fryett2016silicon}, including a new generation of dielectric cavities with extreme confinement of light~\cite{hu2016design,choi2017self,wang2018maximizing}. Our analysis shows that by breaking the translation symmetry, the interaction with the electromagnetic \pk{field} of the resonator leads to the formation of
a localised exciton state with a center-of-mass wave function exactly matching the electromagnetic \pk{field} profile. We find that this localised exciton state can be formally described as an excitonic \emph{reaction coordinate} - a very successful theoretical concept developed in the context of open quantum systems~\cite{garg1985effect,thoss2001self,hughes2009effective,roden2012accounting,iles2014environmental,iles2016energy,martinazzo2011communication}. With this formalism, we derive analytical results for the coupling strength between the exciton and the resonant electromagnetic field as well as the nonlinear exciton-exciton interaction. These results show explicitly that the coupling strength does not scale with the effective mode volume $V$ or the number of excitons $N$. 
Rather, the coupling strength is independent of the lateral \pk{field} 
confinement $L$ and depends only 
on the confinement in the out-of-plane direction $L_z$. The independence of $L$ arises \pk{from a} 
perfect spatial overlap between the exciton reaction coordinate and the \pk{resonant electromagnetic field} 
within the plane.

The representation of the excitons in terms of a single reaction coordinate comes at the price of introducing a residual environment of exciton \pk{states} 
that are coupled to the reaction coordinate, but not directly to the electromagnetic field. 
We show, however, that these residual \pk{exciton states} 
influence the dynamics of the system only at very small lateral confinement length scales\pk{, typically} below a few nanometers.
Within the same reaction coordinate formalism, one can also conveniently account for nonlinear exciton-exciton interactions\pk{, and we} 
find that the lateral \pk{field} 
confinement here plays a crucial role: It determines the effective area of the exciton reaction coordinate, and the nonlinear interaction of excitons within the reaction coordinate therefore scales as $1/L^2$, reflecting \pk{the fact} that the optical confinement dictates the multi-excitonic co-localisation.

We present three different approaches for calculating the time evolution of the system and use these methods to assess the influence of the residual exciton environment and the nonlinear response. In this way, we identify a range of lateral confinement lengths where 
$L$ is sufficiently large that the residual excitons can be ignored but at the same time small enough that the nonlinear response significantly alters the dynamics.

The article is organised as follows. In Sec.~\ref{sec:general-framework}, we describe the Wannier-Mott exciton states \pk{of the 2D materials, the electromagnetic fields of the resonators, and their interaction.} 
In Sec.~\ref{sec:reaction-coordinate}, we derive the exciton reaction coordinate formulation for coupling to a single mode of the electromagnetic resonator. In Sec.~\ref{sec:time-evolution}, we calculate the time evolution of the system using three different approaches, which we benchmark against each other to assess their regimes of validity. In Sec.~\ref{sec:benchm-with-semicl}, we derive the effective linear dielectric function of the Wannier-Mott excitons and use this function to make a reference calculation of the excitation spectrum, which we compare to the microscopic 
quantum reaction coordinate approach in the linear-response limit. Finally, we summarise our findings in Sec.~\ref{sec:conclusion}.


\section{General framework}
\label{sec:general-framework}
In this section, we present the fundamental structure of the theory, which is based on Wannier-Mott exciton states and their interactions with a resonant electromagnetic \pk{field described by a single quasi-normal mode (QNM)}. We will generally study excitons in direct-bandgap semiconductors with discrete in-plane translational symmetry. In this sense, monolayer two-dimensional semiconductors such as transition metal dichalcogenides share many physical features with semiconductor quantum wells, although the excitons of the former are often more strongly bound due to their reduced dielectric screening~\cite{wang2018colloquium}. Monolayer transition-metal dichalcogenides have direct bandgaps at the $K$ and $K'$ points~\cite{cao2012valley,xiao2012coupled} and feature a rich electronic band structure (cf. Fig.~\ref{fig:band-structure}a,c). \pk{In the vicinity of these $K$ and $K'$ points, however, the conduction and valence bands can be well approximated by parabolic bands, leading to an} 
effective-mass approximation, which we shall use here. The optically bright excitons generated from these bands are Coulomb-bound electron-hole states. Comprehensive theoretical treatments of excitons resolve their composite fermionic electron-hole structure~\cite{axt1994dynamics,schafer1996femtosecond,schumacher2006coherent,schafer2013semiconductor,katsch2018theory,katsch2019theory,katsch2020exciton,erkensten2021exciton}. Here, we shall employ a simpler description of the excitons in terms of interacting bosons~\cite{usui1960excitations,marumori1964anharmonic,hanamura1970theory,janssen1971boson,steyn1983quantum}.

The resonant electromagnetic fields in optical cavities and plasmonic particles share many characteristics with the bound states of electrons. 
It is a distinct feature of electromagnetic resonators, however, that the modes are not truly bound, and this gives rise to discrete peaks with finite widths in scattering spectra, for example. From a mathematical point of view, it is advantageous to treat these resonances as modes of the electromagnetic field with finite lifetimes, and the theory of QNMs 
provides a rigorous framework for doing this~\cite{Ching_RevModPhys_70_1545_1998, kristensen2014modes, Lalanne_LPR_12_1700113_2018, Kristensen_AOP_12_612_2020}. \pk{In this work, we will start from a quantum description of the electromagnetic field in electromagnetic resonators~\cite{franke2019quantization} and extend the theory to describe the interaction with Wannier-Mott excitons in 2D materials. At positions far from the resonator, a QNM description of the electromagnetic field is non-trivial, and this poses a challenge for coupling to excitons in the nominally infinite sheet of 2D material. Such a description is in principle influenced by retardation effects, but for the present purpose of describing interactions very close to the resonator we avoid these complications by treating the interaction in the quasi-static limit.}



\subsection{Exciton states}
\label{sec:exciton-states}
\begin{figure}
  \centering
  \includegraphics[width=\columnwidth]{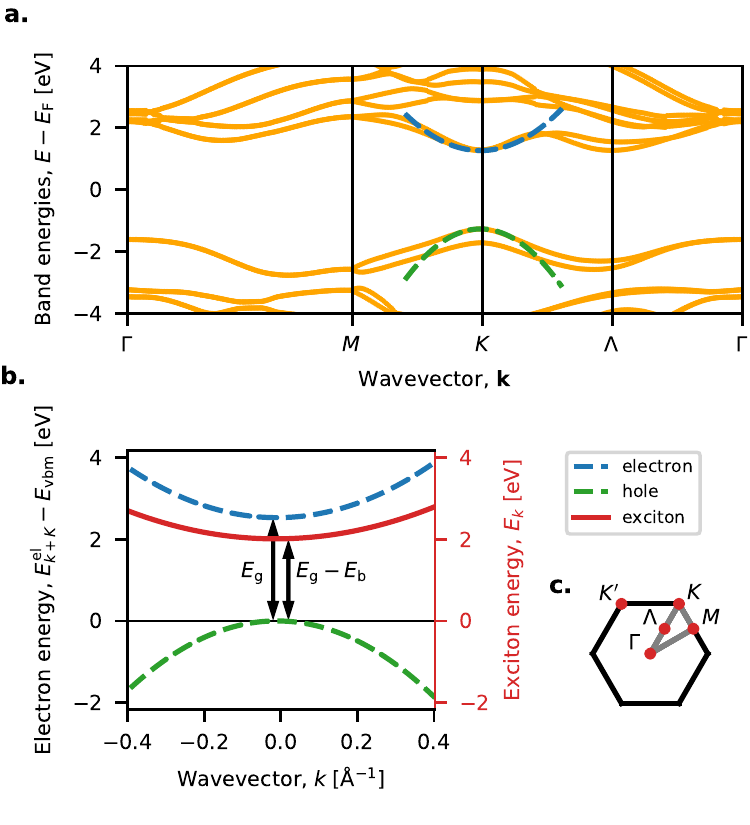}
  \caption{{\bf a.} Electronic band structure (yellow) for monolayer $\mathrm{WS_2}$ (data from $G_0W_0$-calculations of Ref.~\onlinecite{haastrup2018computational}). The band energies are plotted \rev{as function of} 
in-plane wave vector and are given relative to the Fermi energy, $E_{\rm F}$. Within the effective-mass approximation, the conduction and valence band energies near the band gap, ($K$ or $K'$), are approximated as parabolic (blue and green dashed lines). {\bf b.} Conduction band (blue) and valence band (green) band electronic energies \revv{relative to valence band maximum, $E_{\rm vbm}$ (left axis)} as a function of the wavevector, $\mathbf{k}$, given relative to $K$ (or $K'$). Also shown is the lowest-lying 1s exciton energy \revv{(right axis)} as a function of the center-of-mass momentum, $\mathbf{k}$. The band gap, $E_{\rm g}$ and the exciton gap, $E_0=E_{\rm g}-E_{\rm b}$, are indicated with vertical arrows. \revv{The effective masses of the valence and conduction bands are independent of the direction in k-space~\cite{haastrup2018computational} and the exciton dispersion is thus direction-independent as well.}  {\bf c.} The Brillouin zone of a 2D hexagonal lattice with special symmetry points marked as red dots and the path over which the band structure in panel a is calculated (grey line).}
  \label{fig:band-structure}
\end{figure}

The description of the excitonic degrees of freedom in the 2D material will be based on a Wannier-Mott framework~\cite{grosso2013solid}, which provides a useful analytical description of the excitons and has proven successful in this context~\cite{latini2015excitons, olsen2016simple}. To this end, we consider the lowest-energy conduction band and the highest-energy valence band of a two-dimensional semiconductor, separated by a band gap energy $E_{\rm g}$. We denote by $\ket{0}$ the Fermi sea, i.e. the fermionic state in which the valence band is fully occupied by electrons and the conduction band is empty. The fermionic creation operator for a hole in the valence band with in-plane wavevector $\mathbf{k}$ is denoted 
by $v^\dagger_{\alpha\mathbf{k}}$ and, similarly, $c_{\alpha\mathbf{k}}^\dagger$ denotes the creation operator for an electron in the conduction band. Here, the index $\alpha$ labels the high-symmetry point with wavevector $\mathbf{K}_\alpha$, when the band gap is degenerate, as is the case for e.g. monolayer transition-metal dichalcogenides, where $\alpha\in\{K,K'\}$~\cite{wang2018colloquium}. The single-particle wave functions, $\psi_{\mathrm{c},\alpha\mathbf{k}}(\mathbf{r})$ and $\psi_{\mathrm{v},\alpha\mathbf{k}}(\mathbf{r})$ are taken to be of Bloch form,
\begin{align}
\label{eq:bloch-wave functions}
  \psi_{i,\alpha\mathbf{k}}(\mathbf{r}) = \frac{1}{\sqrt{N}} e^{\text{i}(\mathbf{k}+\mathbf{K}_\alpha)\cdot\mathbf{r}} u_{i,\alpha}(\mathbf{r}),
\end{align}
where $i=\mathrm{c,v}$; $N$ is the number of unit cells in the 2D sheet with surface area $S$, and $u_{i,\alpha}(\mathbf{r})$ is a Bloch function, which has the periodicity of the crystal lattice and is normalised over a single unit cell, $\int_{V_{\rm UC}}\dd[3]{\mathbf{r}}\abs{u_{i,\alpha}(\mathbf{r})}^2=1$. Within the Wannier-Mott framework, the conduction band and valence band energies are described through effective masses, $m_\text{e}$ and $m_\text{h}$, respectively, which reflect the local parabolic approximation to the band structure near the $K$ and $K'$ valleys, cf. Fig.~\ref{fig:band-structure}a.


The Wannier-Mott exciton states can be written as a momentum-superposition of electron-hole pairs~\cite{savona1994quantum}
\begin{align}
\label{eq:single-exciton-state}
  \ket{\Phi_{n,\alpha\mathbf{k}}} = \sum_\mathbf{q} \phi_n(\mathbf{q}) 
  \hat{c}_{\alpha,(m_\text{e}/M)\mathbf{k} + \mathbf{q}}^\dagger \hat{v}_{\alpha,(m_\text{h}/M) \mathbf{k} - \mathbf{q}}^\dagger \ket{0},
\end{align}
where $\phi_n(\mathbf{q})$ 
is the momentum-space exciton wave function with shell index $n$, and $M=m_e+m_h$ is the total exciton mass. The label $\mathbf{k}$ is thus the center-of-mass momentum of the exciton. In Ref.~\onlinecite{olsen2016simple}, it was found that a very good approximation to the lowest-energy exciton ($n=1$) for several two-dimensional transition-metal dichalcogenides is given by the simple hydrogenic form
\begin{align}
\label{eq:exciton-phi-q}
  \phi(\mathbf{q}) = \frac{\sqrt{8\pi a_\mathrm{B}^2/S}}{[1+(qa_\mathrm{B})^2]^{3/2}},
\end{align}
in which $a_\mathrm{B}$ is the exciton Bohr radius and $q=\abs*{\mathbf{q}}$. This form is often found in the context of semiconductor quantum wells~\cite{tassone1999exciton,rochat2000excitonic}. In the present work, we restrict our discussion to the lowest-lying exciton and generally use Eq.~\eqref{eq:exciton-phi-q} for the exciton wave function and denote its quantum state by $\ket{\Phi_{\alpha, \bf k}}$, dropping the index $n$ in Eq.~\eqref{eq:single-exciton-state}.
The total exciton energy is given by the sum of the kinetic energy of the center-of-mass coordinate, the band gap $E_{\rm g}$ and the exciton binding energy, $E_{\rm b}$:
\begin{align}
  E_\mathbf{k} = E_0 + \frac{\hbar^2 k^2}{2M},
\end{align}
where $E_0 := E_{\rm g} - E_{\rm b}$ is the exciton gap, as illustrated in Fig.~\ref{fig:band-structure}b. We will often use the corresponding frequencies as well, $\omega_\mathbf{k}=E_\mathbf{k}/\hbar,\;$ and $\omega_0=E_0/\hbar$.

Being composed of electron-hole pairs, it is favourable 
to describe the excitons through a set of bosonic creation operators, $\hat{b}_{\alpha\mathbf{k}}^\dagger$, which \rev{generate} 
single- and multi-excitonic quantum states. Noting, however, that operators of the form $C^\dagger_{\alpha\mathbf{k}}=\sum_\mathbf{q} \phi(\mathbf{q}) \hat{c}_{\alpha,(m_\text{e}/M)\mathbf{k} + \mathbf{q}}^\dagger \hat{v}_{\alpha,(m_\text{h}/M) \mathbf{k} - \mathbf{q}}^\dagger $ have commutation relations that are neither bosonic nor fermionic~\cite{keldysh1968collective}, a description of the excitons as non-interacting bosons is infeasible. Nevertheless, it is possible to use an 
approximate description in terms of interacting bosons~\cite{usui1960excitations,marumori1964anharmonic,hanamura1970theory,janssen1971boson,steyn1983quantum}.
In this work, we include the leading interaction term, which arises due to exciton-exciton Coulomb interactions~\cite{tassone1999exciton,rochat2000excitonic}, such that the total bosonic form of the exciton Hamiltonian (including the free exciton energy) is
\begin{align}
  \begin{split}
\hat{H}_{\rm x} = &\sum_{\alpha\mathbf{k}} \hbar\omega_\mathbf{k}\hat{b}_{\alpha\mathbf{k}}^\dagger \hat{b}_{\alpha\mathbf{k}} \\ &+ \sum_{\alpha\mathbf{k}\mathbf{k}'\mathbf{q}} \hbar W_{\mathbf{k}\mathbf{k}'\mathbf{q}} \hat{b}^\dagger_{\alpha\mathbf{k}+\mathbf{q}}\hat{b}^\dagger_{\alpha\mathbf{k}'-\mathbf{q}}\hat{b}_{\alpha\mathbf{k}'}\hat{b}_{\alpha\mathbf{k}},
\end{split}
\end{align}
where $\hbar W_{\mathbf{k}\mathbf{k}'\mathbf{q}}$ is the momentum-dependent exciton-exciton interaction energy and $\hat{b}_{\alpha\mathbf{k}}$ and $\hat{b}_{\alpha\mathbf{k}}^\dagger$ are bosonic exciton annihilation and creation operators, obeying the commutation relation $[\hat{b}_{\alpha\mathbf{k}},\hat{b}_{\alpha',\mathbf{k}'}^\dagger] = \delta_{\mathbf{k}\mathbf{k}'}\delta_{\alpha\alpha'}$. %
These operators can be understood as a bosonic approximation to the excitonic operators $\hat{C}_{\alpha\mathbf{k}}$ and $\hat{C}^\dagger_{\alpha\mathbf{k}}$, such that $\hat{b}_{\alpha\mathbf{k}}^\dagger\ket{0'}$ is the bosonic representation of $\ket{\Phi_{\alpha\mathbf{k}}}$. The state $\ket{0'}$ is the bosonic exciton vacuum state, which is the bosonic equivalent to the Fermi sea~\cite{usui1960excitations}\pk{. In the present context, the correspondence between excitonic fermion pairs and bosons is only of formal interest, since the practical calculations of the microscopic light-matter coupling strength can be phrased in terms of the fermionic operators. Similarly, the interaction strengths $W_{\mathbf{k}\mathbf{k}'\mathbf{q}}$ are directly related to four-particle scattering matrix elements evaluated in the original fermionic space~\cite{usui1960excitations,hanamura1970theory, tassone1999exciton, rochat2000excitonic}.} %
%
For convenience, we can split the exciton Hamiltonian into a noninteracting part, $\hat{H}_{\rm x,0} = \sum_{\alpha\mathbf{k}} \hbar\omega_\mathbf{k}\hat{b}_{\alpha\mathbf{k}}^\dagger \hat{b}_{\alpha\mathbf{k}}$, and an interacting part, $\hat{W} = \sum_{\alpha\mathbf{k}\mathbf{k}'\mathbf{q}} \hbar W_{\mathbf{k}\mathbf{k}'\mathbf{q}} \hat{b}^\dagger_{\alpha\mathbf{k}+\mathbf{q}}\hat{b}^\dagger_{\alpha\mathbf{k}'-\mathbf{q}}\hat{b}_{\alpha\mathbf{k}'}\hat{b}_{\alpha\mathbf{k}}$. Detailed discussions and calculations of the interaction strengths $W_{\mathbf{k}\mathbf{k}'\mathbf{q}}$ can be found in 
Refs.~\cite{tassone1999exciton,rochat2000excitonic,ciuti1998role,shahnazaryan2017exciton}. In general, a momentum cutoff of the order $a_{\rm B}^{-1}$ is reported, which means 
that $\hbar W_{\mathbf{k}\mathbf{k}'\mathbf{q}}\simeq \hbar W_{000}$ for $k,k',q \ll a_{\rm B}^{-1}$. In Ref.~\onlinecite{shahnazaryan2017exciton}, variational calculations of the exciton wavefunctions and the corresponding Coulomb matrix elements showed that for monolayer $\mathrm{WS_2}$, the interaction strength can be approximated as $\hbar W_{000}\simeq 2.07 E_{\rm b} a_{\rm B}^2/S$. It was also shown that the interaction has contributions from a direct part, where the constituent electron and hole within the two interacting excitons remain fixed, and an exchange part, where the constituent particles are exchanged. In particular, it was shown that the exchange part dominates at low momenta, and that the direct part vanishes identically at zero momentum, meaning that the matrix element $W_{000}$ is determined solely by the exchange contribution. We note, furthermore, that intervalley exchange Coulomb effects in transition-metal dichalcogenides~\cite{qiu2015nonanalyticity} are not included in this work. Besides from an overall shift, the linear exchange coupling vanishes for vanishing exciton center-of-mass momentum~\cite{qiu2015nonanalyticity,deilmann2019finite,guo2019exchange}, and is therefore small for optically bright excitons, which have small center-of-mass momenta. The nonlinear intervalley exchange Coulomb interaction is small compared to the intravalley Coulomb interaction, $\hat{W}$~\cite{katsch2020exciton}. Thus, we expect that the intervalley exchange interaction will give rise to minor corrections to the overall physics of light-matter coupling and the weak nonlinear optical response studied here.

\begin{table*}
  \centering
  \begin{tabular}{l|c|c|c|c|l|c}
    Material & $m_{\rm h} \;[m_0]$ & $m_{\rm e} \; [m_0]$ & $E_{\rm g} \; [\mathrm{eV}]$ & $ E_{\rm b} \; [\mathrm{eV}]$ & $a_{\rm B} \; [\mathrm{nm}]$ & $\mathscr{V}\; [\mathrm{m/s}]$
    \\[.25mm] \hline
    $\mathrm{WS_2}$ \phantom{$\hat{\hat{\text{I}}}$} & 0.34 \cite{haastrup2018computational} & 0.33 \cite{haastrup2018computational}& 2.53 \cite{haastrup2018computational} & 0.52 \cite{haastrup2018computational} & 1.95 \cite{li2014measurement} & $6.7\times 10^5$~~\cite{xiao2012coupled}
    \\[.25mm]
    $\mathrm{MoS_2}$ & 0.53 \cite{haastrup2018computational}  & 0.43 \cite{haastrup2018computational} & 2.53 \cite{haastrup2018computational} & 0.55 \cite{haastrup2018computational} & 2.0~\cite{li2014measurement} 1.68~\cite{zhang2014absorption} &  $5.3\times 10^{5}$~~\cite{xiao2012coupled}
    \\[.25mm]
$\mathrm{WSe_2}$ & 0.36 \cite{haastrup2018computational} & 0.39 \cite{haastrup2018computational} & 2.10 \cite{haastrup2018computational} & 0.48 \cite{haastrup2018computational} & 3.3~\cite{li2014measurement} & $6.0\times 10^5$~~\cite{xiao2012coupled}
    \\[.25mm]
$\mathrm{MoSe_2}$ & 0.58 \cite{haastrup2018computational} & 0.49 \cite{haastrup2018computational}& 2.12 \cite{haastrup2018computational} & 0.50 \cite{haastrup2018computational} & 2.6~\cite{li2014measurement} & $4.7\times 10^5$~~\cite{xiao2012coupled}
  \end{tabular}
  \caption{Relevant material parameters with referenced sources for a selection of monolayer transition metal dichalcogenides. The listed quantities are: Effective electron and hole masses, $m_{\rm e}$ and $m_{\rm h}$, respectively, given in units of the free electron mass, $m_0$; the bandgap energy, $E_{\rm g}$; the exciton binding energy, $E_{\rm b}$; the exciton Bohr radius, $a_{\rm B}$; and the velocity parameter, $\mathscr{V}$, used for calculation of the Bloch momentum matrix elements. The band gaps from Ref.~\onlinecite{haastrup2018computational} have been taken from $G_0W_0$ band-structure calculations. The exciton Bohr radius \rev{has} been calculated using the oscillator strength from the experimentally measured dielectric response of monolayers exfoliated on fused silica substrates in Ref.~\onlinecite{li2014measurement}, as described in Sec.~\ref{sec:benchm-with-semicl}. For $\mathrm{WS_2}$, the oscillator strength is explicitly given in Ref.~\onlinecite{li2014measurement}. For the other materials, we have fitted six Lorentzian oscillators to the data for the imaginary part of the dielectric function in Ref.~\onlinecite{li2014measurement} to extract the oscillator strength. A similar calculation based on absorption measurements was made in Ref.~\onlinecite{zhang2014absorption} for $\mathrm{MoS_2}$.}
  \label{tab:2D-mat-parameters}
\end{table*}

A collection of relevant material parameters for a selection of monolayer transition-metal dichalcogenides is presented in Table~\ref{tab:2D-mat-parameters}.


\subsection{Resonant electromagnetic fields}
\label{sec:resonant-electromagnetic-fields}
Optical cavities and plasmonic particles support a number of resonances, which show up as distinct peaks in scattering spectra, for example, and whose corresponding field distributions are commonly referred to as quasi-normal modes (QNMs)~\cite{Ching_RevModPhys_70_1545_1998, kristensen2014modes,Lalanne_LPR_12_1700113_2018, Kristensen_AOP_12_612_2020} or resonant states~\cite{Muljarov_EPL_92_50010_2010,Muljarov_OL_43_1978_2018}. They are defined as solutions to the wave equation subject to suitable radiation conditions, such as the Silver-M{\"u}ller condition for resonators in free space~\cite{martin2006multiple,kristensen2015normalization}. For the present application, we limit the analysis to cases in which there is only a single QNM in the frequency range of interest, and we denote the associated electric-field distribution by $\mft_\text{c}(\mr)$; the corresponding eigenfrequency $\tlo_{\text{c}} = \omega_\text{c} - \mathrm{i}\gamma_\text{c}$ is complex with a negative imaginary part, where $\gamma_{\rm c}$ accounts for the cavity decay rate. 
For positions close to or inside the resonator, the operator describing the quantized electric field can be expanded in terms of this QNM by the method presented in Ref.~\onlinecite{franke2019quantization}.
%
In this work, we shall be interested in coupling to sheets of 2D materials, which extend to regions far away from the resonator, and where the formulation of Ref.~\onlinecite{franke2019quantization} is not directly applicable. As discussed in Appendix~\ref{App:Electric_field_outside_the_resonator}, however, it is possible to extend the general framework of Ref.~\onlinecite{franke2019quantization} by use of the Lippmann-Schwinger equation and ideas originally put forward in Ref.~\onlinecite{Ge_NJP_16_113048_2014} to write the electric-field operator at general positions $\mr$ outside the resonator in terms of a convolution as
\begin{align}
\mEh(\mr,t) = \text{i}\sqrt{\frac{\hbar\omega_\text{c}}{2\epsilon_0}}\int_0^\infty\dd{\tau}\mFt_\text{c}(\mr,t-\tau)\hat{a}_\text{c}(\tau) + \text{H.c.}
\label{Eq:mEh_time_retarded_convolution}
\end{align}
Here, the creation and annihilation operators $\hat{a}_\text{c}^\dagger$ and $\hat{a}_\text{c}$ obey the commutation relation $[\hat{a}_\text{c}(t),\hat{a}_\text{c}^\dagger(t)]=1$~\cite{franke2019quantization}, and the memory kernel $\mFt_\text{c}(\mr,t)$, 
which ensures a proper causal relation to locations far from the resonator, derives from analytical continuation of the electric field QNM onto the real frequency axis. In the present work, we shall focus on the local and non-retarded coupling dynamics by setting $\mFt_\text{c}(\mr,t)=\mFt_\text{c}(\mr)\delta(t)$ and calculating $\mFt_\text{c}(\mr)$ from $\mft_\text{c}(\mr)$ by use of the quasistatic Green tensor, see Appendix~\ref{App:Electric_field_outside_the_resonator} for details. Moreover, as we shall see below, the coupling will be phrased in terms of the electromagnetic vector potential, which we write in the single-QNM case as
\begin{align}
\hat{\mathbf{A}}(\mathbf{r}) = \sqrt{\frac{\hbar}{2\epsilon_0\omega_\text{c}}}  [\hat{a}_\text{c} \mFt_\text{c}(\mathbf{r}) + \hat{a}_\text{c}^\dagger\mFt_{\rm c}^*(\mathbf{r})].
\label{Eq:A_single_QNM_expansion}
\end{align}

Due to radiative loss and possibly absorption in the resonator material, the energy of the electromagnetic field in the resonator is not conserved. Therefore, it is convenient to describe the state of the field through its density operator, $\hat{\rho}_{\rm c}$, which in the absence of interactions is governed by the master equation~\cite{franke2019quantization}
\begin{align}
\dv{\hat{\rho}_\text{c}}{t} = -\text{i}[\omega_{\rm c} \hat{a}_{\rm c}^\dagger  \hat{a}_{\rm c} ,\hat{\rho}_\text{c}] + 2\gamma_\text{c}\mathcal{D}(\hat{a}_\text{c},\hat{\rho}_\text{c}),
\end{align}
where
\begin{align}
  \label{eq:lindblad-dissipator}
\mathcal{D}(\hat{x},\hat{\rho}) = \hat{x}\hat{\rho} \hat{x}^\dagger -\frac{1}{2}(\hat{x}^\dagger \hat{x} \hat{\rho} + \hat{\rho} \hat{x}^\dagger \hat{x})
\end{align}
is the Lindblad dissipator.

\subsection{Light--matter interaction }
\label{sec:light-matt-inter}
We now proceed to consider interactions between the excitons and the resonant electromagnetic field. The exciton-field coupling is generated by the minimal-coupling Hamiltonian, of which the dominating part (written in first quantization) is~\cite{girlanda1981two}
\begin{align}
\hat{H}_{\rm I}  = -\frac{e_0}{m_0}\sum_i \hat{\mathbf{A}}(\mathbf{r}_i)\cdot\hat{\mathbf{p}}_i,
\end{align}
where the sum runs over the electrons in the system.

The light--matter interaction Hamiltonian can be cast into the second-quantised form~\cite{savasta1995particle}
\begin{align}
\label{eq:H-light-matter-2nd-quant}
\hat{H}_{\rm I} = \hbar\sum_{\alpha\mathbf{k}}\qty[ \hat{b}_{\alpha\mathbf{k}}^\dagger(g_{\alpha\mathbf{k}}\hat{a}_\text{c} + g_{\alpha\mathbf{k}}' \hat{a}_\text{c}^\dagger)
 + \hat{b}_{\alpha\mathbf{k}}(g_{\alpha\mathbf{k}}^{\prime *}\hat{a}_\text{c} + g_{\alpha\mathbf{k}}^{*} \hat{a}_{\text{c}}^\dagger)],
\end{align}
where the interaction strengths are evaluated in the fermionic space as
\begin{align}
\hbar g_{\alpha\mathbf{k}} = -\frac{e_0}{m_0}\sqrt{\frac{\hbar}{2\epsilon_0\omega_{\text{c}}}}\mel{\Phi_{\alpha\mathbf{k}}}{\sum_i\mFt_{\text{c}}(\mathbf{r_i})\cdot\hat{\mathbf{p}}_i}{0},
\end{align}
and $\hbar g_{\alpha\mathbf{k}}'$ is obtained by replacing ${\mFt}_{\text{c}}(\mathbf{r})$ with $\mFt_\text{c}^*(\mathbf{r})$.
Assuming that the light--matter coupling is weak compared to the relevant electromagnetic frequencies and exciton frequencies, we make the rotating-wave approximation, corresponding to setting $g_{\alpha\mathbf{k}}'=0$ in Eq.~\eqref{eq:H-light-matter-2nd-quant}. With these approximations, the light--matter interaction is
\begin{align}
\label{eq:light-matter-interaction-RWA-single-mode}
\hat{H}_{\rm I} = \sum_{\alpha\mathbf{k}}\qty(\hbar g_{\alpha\mathbf{k}}\hat{a}_{\rm c}\hat{b}_{\alpha\mathbf{k}}^\dagger + \hbar g_{\alpha\mathbf{k}}^*\hat{a}_{\rm c}^\dagger \hat{b}_{\alpha\mathbf{k}}).
\end{align}

Using the Slater-Condon rules~\cite{grosso2013solid}, the matrix element entering $g_{\alpha\mathbf{k}}$ can be evaluated, again in the fermionic space, as
\begin{align}
\label{eq:slater-matrix-element-light-matter}
\begin{split}
&\mel{\Phi_{\alpha\mathbf{k}}}{\sum_i{\mFt}_{\text{c}}(\mathbf{r}_i)\cdot\hat{\mathbf{p}}_i}{0} = \\ &\hspace{1.5cm}\frac{1}{N}\sum_\mathbf{q}{\phi(\mathbf{q})}\int\dd[3]{\mathbf{r}}e^{-i\mathbf{k}\cdot\mathbf{r}} u_{\rm c, \alpha}^*(\mathbf{r}){\hat{\mathbf{p}}}\cdot{\mFt}_{\text{c}}(\mathbf{r})u_{\rm v,\alpha}(\mathbf{r}),
\end{split}
\end{align}
{where we used the expression for the single-particle wave functions of Eq.~(\ref{eq:bloch-wave functions}).}
The summation over $\mathbf{q}$ only involves the wave function ${\phi(\mathbf{q})}$ and can thus be evaluated independently as
\begin{align}
  \sum_\mathbf{q} {\phi(\mathbf{q})} = \frac{S}{(2\pi)^2} \int \dd[2]{\mathbf{q}} {\phi(\mathbf{q})}
= \sqrt{\frac{2 S}{\pi a_{\rm B}^2}}.
\end{align}
Assuming that the mode function does not vary appreciably over the unit cell, and that the relevant wavevectors are much smaller than the inverse lattice constant, we can approximate the integral in Eq.~\eqref{eq:slater-matrix-element-light-matter} as a sum over unit cells (indexed by $j$) and a Bloch matrix element
\begin{align}
  \label{eq:integral-to-sum-unit-cell}
  \begin{split}
\int\dd[3]{\mathbf{r}}e^{-\mathrm{i}\mathbf{k}\cdot\mathbf{r}} &u_{\rm c}^*(\mathbf{r})\mathbf{p}\cdot{\mFt}_{\text{c},\alpha}(\mathbf{r})u_{\rm v,\alpha}(\mathbf{r}) \\
&= \sum_j e^{-\mathrm{i}\mathbf{k}\cdot\mathbf{r}_j}{\mFt}_{\text{c}}(\mathbf{r}_j,z_0)\cdot \mathbf{p}_\mathrm{cv}^\alpha,
\end{split}
\end{align}
where $\mathbf{p}_\mathrm{cv}^\alpha = \int_{V_{\rm UC}}\dd[3]{\mathbf{r}} u_{\rm c,\alpha}^*(\mathbf{r})\hat{\mathbf{p}}u_{\rm v,\alpha}(\mathbf{r})$, $\mathbf{r}_j$ is the lateral coordinate of the $j$th unit cell, and we have taken the 2D-sheet to be located at $z=z_0$.
For 2D materials in the transition-metal dichalcogenide family, the two degenerate exciton modes at the $K$ and $K'$ valleys have the matrix elements~\cite{xiao2012coupled,wang2016radiative}, $\mathbf{p}_\mathrm{cv}^K = m_0\mathscr{V}(\mathbf{x} +{\text{i}}\mathbf{y}), \; \mathbf{p}_\mathrm{cv}^{K'} = m_0\mathscr{V}(\mathbf{x} -{\text{i}}\mathbf{y})$, where $\mathscr{V}$ is a material-dependent velocity parameter (see Table~\ref{tab:2D-mat-parameters} for values). These matrix elements are circularly polarized due to spin-orbit coupling.
The summation over $j$ in Eq.~\eqref{eq:integral-to-sum-unit-cell} can then be rewritten as an integral as $\sum_j\rightarrow \frac{N}{S}\int\dd[2]{\mathbf{r}}$. Combining everything, the coupling strength becomes
\begin{align}
\label{eq:g-k-p.A}
\hbar g_{\alpha\mathbf{k}} = -\frac{e_0}{m_0}\sqrt{\frac{\hbar}{\pi\epsilon_0\omega_{\text{c}} a_{\rm B}^2 S}}\int\dd[2]{\mathbf{r}}e^{-\mathrm{i}\mathbf{k}\cdot\mathbf{r}}{\mFt}_{\text{c}}(\mathbf{r},z_0)\cdot\mathbf{p}_\mathrm{cv}^\alpha,
\end{align}
where the integral is over the infinite extent of the 2D material. \pk{Whereas the QNMs diverge (exponentially) at sufficiently large distances from the resonator~\cite{Kristensen_AOP_12_612_2020}, the functions $\mFt_\text{c}(\mr)$ behave as the free-space quasistatic Green tensor at sufficiently large distances and are therefore square integrable.}

We note that in addition to the contributions to the light-matter interactions derived until this point, a nonlinear saturation term of the form
\begin{align}
  \begin{split}
\hat{W}_\text{EM}=&\sum_{\alpha\mathbf{k}_1\mathbf{k}_2\mathbf{k}_3}\hbar\sigma_{\alpha\mathbf{k}_1\mathbf{k}_2\mathbf{k}_3}(\hat{b}_{\alpha\mathbf{k}_1}^\dagger \hat{b}_{\alpha\mathbf{k}_2}\hat{b}_{\alpha\mathbf{k}_3} + \mathrm{H.c.})\\ &\times (\hat{a}_{\text{c}}^\dagger + \hat{a}_{\text{c}})
\end{split}
\end{align}
appears due {to} the non-bosonic nature of the excitons~\cite{tassone1999exciton,rochat2000excitonic}. Within the hydrogenic Wannier-Mott exciton approximation, the nonlinear interaction due to saturation can be approximated in the zero-momentum limit, where it takes its maximal value, as $\hbar\sigma_{\alpha 000}\simeq (4\pi/7) (a_{\rm B}^2/S) 2G_0$, where $G_0$ is the coupling strength~\cite{tassone1999exciton}. For a coupling strength of 50 meV and an exciton binding energy of $500$ meV, this gives a ratio relative to the exciton-exciton Coulomb interaction of $\sigma_{\alpha 000}/W_{000}\simeq 0.17$. Thus, for the systems of interest in this article, the Coulomb exciton-exciton interaction $\hat{W}$ is considered to be the strongest nonlinear effect, and the saturation interaction is neglected.
We do note, however, that such saturation effects have been shown to play an important role for the nonlinear dynamics of trions in transition-metal dichalcogenides, owing to the three-fermion composite structure of trions~\revv{\cite{kyriienko2020nonlinear,emmanuele2020highly}}.

\section{Exciton reaction coordinate}
\label{sec:reaction-coordinate}
In Section~\ref{sec:general-framework}, we have seen that the excitons are described by a Hamiltonian $\hat{H}_{\rm x}$, which conserves the center-of-mass momentum $\mathbf{k}$, thus reflecting the discrete translation symmetry of the 2D material sheet. This symmetry is then explicitly broken through the interaction Hamiltonian $\hat{H}_{\rm I}$, because the electric-field distribution $\tilde{\mathbf{F}}_{\rm c}$ is not translationally invariant. In this section, we introduce a basis transformation of the excitons that provides a natural starting point for analyzing this system. The transformation defines a localized exciton mode, which we denote the \emph{exciton reaction coordinate} in the spirit of quantum chemistry, where the concept of reaction coordinates has been developed in a similar fashion to describe nuclear motion in molecules~\cite{garg1985effect,thoss2001self,hughes2009effective,roden2012accounting,iles2014environmental,iles2016energy}. In the new basis, the light-matter interaction is greatly simplified, because the electromagnetic field couples only to the exciton reaction coordinate. The transformation comes at the cost of introducing a reservoir of residual exciton modes, which in turn are coupled to the reaction coordinate. However, as we shall see in Section~\ref{sec:time-evolution}, there are several successful approximate strategies for treating these residual modes.

\subsection{Exciton reaction coordinate and residual excitonic spectral density}
\label{sec:single-electr-mode}

We now define a new basis of exciton modes, $\{\hat{B}_i\}$, generated by a unitary transformation $U$ as~\cite{martinazzo2011communication}
\begin{align}
\hat{B}_i = \sum_{\alpha\mathbf{k}} U_{i\alpha\mathbf{k}} \hat{b}_{\alpha\mathbf{k}},
\end{align}
{in which the first row in the} 
transformation matrix 
{$U$} 
is given by $U_{0\alpha\mathbf{k}}=g_{\alpha\mathbf{k}}^*(\sum_{\alpha\mathbf{k}}\abs*{g_{\alpha\mathbf{k}}}^2)^{-1/2}$, such that
\begin{align}
\hat{B}_0 = \Big[\sum_{\alpha\mathbf{k}} \abs*{g_{\alpha\mathbf{k}}}^2 \Big]^{-1/2}\sum_{\alpha\mathbf{k}} g_{\alpha\mathbf{k}}^* \hat{b}_{\alpha\mathbf{k}}.
\end{align}
We denote this collective mode as the \emph{exciton reaction coordinate}.
The remaining rows in $U$ are constructed {via Gram-Schmidt orthogonalisation} as orthonormal vectors{, such that} 
$U$ is unitary, i.e.
\begin{align}
  \sum_{\alpha\mathbf{k}} U^*_{j\alpha\mathbf{k}} U_{i\alpha\mathbf{k}}  = \delta_{ij}.
\end{align}
The light--matter interaction can now be written in the much simpler form
\begin{align}
\hat{H}_{\rm I} = \hbar G_0(\hat{B}_0 \hat{a}^\dagger + \hat{B}_0^\dagger \hat{a}),
\end{align}
where $G_0 = \sqrt{\sum_{\alpha\mathbf{k}} \abs*{g_{\alpha\mathbf{k}}}^2}$. This
illustrates the advantage of introducing the reaction coordinate: by construction, the resonant electromagnetic field now only interacts with the weighted sum of Wannier-Mott excitons defined by $\hat{B}_0$. {The transformation $U$ thus precisely captures the notion of a localized excitonic state, which is defined by the interaction with the electromagnetic field. This physically appealing reformulation of the dynamics comes at the computational price that the associated} 
free-exciton Hamiltonian {$\hat{H}_{x,0}$} is no longer diagonal{,}
{
\begin{align}
    \hat{H}_{x,0} = \sum_{ii'}\hbar\Omega_{ii'}\hat{B}_i^\dagger \hat{B}_{i'},
\end{align}
where
$\Omega_{ii'} = \sum_{\alpha\mathbf{k}}E_\mathbf{k} U_{i\alpha\mathbf{k}}U_{i'\alpha\mathbf{k}}^*$ and thus $\Omega_{ii'} = \Omega_{i'i}^*$.
} 
In this way, the exciton reaction coordinate is coupled to a \emph{residual environment} of other exciton modes, $\hat{B}_{i}, \:i>0$. As we shall see later, it is useful to have access to the spectral density of this environment, {and this 
requires a re-diagonalisation as follows. The} 
procedure starts by separating out the terms in $\hat{H}_{\rm x,0}$ that contain the reaction coordinate 
{$\hat{B}_0$},
\begin{align}
  \label{eq:H-exc-RC-1}
  \begin{split}
    \hat{H}_{\rm x,0} = &\hbar\Omega_{00} \hat{B}_0^\dagger \hat{B}_0 + \hbar\sum_{i>0} \qty(\Omega_{0i} \hat{B}_0^\dagger \hat{B}_i + \Omega_{i0}\hat{B}_i^\dagger \hat{B}_0) \\ &+ \sum_{ii'>0} \hbar\Omega_{ii'}\hat{B}_i^\dagger \hat{B}_{i'}{.}
  \end{split}
\end{align}
This can be written in the compact matrix form
\begin{align}
   \hat{H}_{\rm x,0} = \hbar\Omega_{00} \hat{B}_0^\dagger \hat{B}_0 + \hat{\mathbf{B}}^\dagger \hbar {\boldsymbol{\lambda}} \hat{B}_0 + \hat{B}_0\hbar {\boldsymbol{\lambda}}^\dagger\hat{\mathbf{B}}  +  \hat{\mathbf{B}}^\dagger \hbar\Omega' \hat{\mathbf{B}},
  \label{eq:H-exc-RC-1_matrix_form}
\end{align}
where $\Omega'$ is {constructed from $\Omega$ by removing the} 
first row and column, and {$\hat{\mathbf{B}}$ and $\boldsymbol{\lambda}$ are vectors} with elements {$\hat{B}_i$} and $\lambda_i = \Omega_{i0}${, respectively, with $i>0$}.
 We now wish to re-diagonalise the part of the Hamiltonian governing the modes with $i>0$. To do this, we define $\tilde{U}$ as the transformation that diagonalises $\Omega'$, such that $\tilde{\Omega}:=\tilde{U}^\dagger\Omega'\tilde{U}$ is diagonal. We also define a transformed set of modes, $\hat{\mathbf{B}} = \tilde{U}\hat{\tilde{\mathbf{B}}}$.
Since {$\Omega'$} is a Hermitian matrix, $\tilde{U}$ is {unitary,} 
and the columns of $\tilde{U}$ are the eigenvectors of {$\Omega'$. In this way, the last term in Eq.~(\ref{eq:H-exc-RC-1_matrix_form}) can be written in a diagonal form, and the 
free-exciton Hamiltonian can be written as}
\begin{align}
\label{eq:H-exc-RC-2}
\hat{H}_{\rm x,0} = \hbar\Omega_0 \hat{B}_0^\dagger \hat{B}_0 + \sum_{i>0} \hbar\tilde{\Omega}_i \hat{\tilde{B}}_i^\dagger \hat{\tilde{B}}_i + \hbar\tilde{\lambda}_i\hat{B}_0\hat{\tilde{B}}_i^\dagger + \hbar\tilde{\lambda}_i^*\hat{B}_0^\dagger \hat{\tilde{B}}_i ,
\end{align}
where $\tilde{\lambda} = \tilde{U}^\dagger\lambda$ and we have defined $\Omega_0:=\Omega_{00},\; \tilde{\Omega}_i := \tilde{\Omega}_{ii}$. {From Eq.~(\ref{eq:H-exc-RC-2}) it is evident that the exciton states of the residual environment 
are mutually uncoupled and couple only to the exciton reaction coordinate.} The spectral density of the residual environment can be calculated from Eq.~\eqref{eq:H-exc-RC-2} as
\begin{align}
J_{\rm res}(\omega) = \sum_{i>0}\abs*{\tilde{\lambda}_i}^2\delta(\omega-\tilde{\Omega}_{i}).
\end{align}
This is in short called the \emph{residual spectral density} of the exciton reaction coordinate. Although the procedure of constructing the reaction coordinate transformation matrix {$U$} and subsequently the re-diagonalisation matrix $\tilde{U}$ can be carried out numerically, this scales poorly with the number of {elements in the matrix}.  {Fortunately,} 
the procedure of extracting reaction coordinates and residual spectral densities has been extensively studied in the literature of open quantum systems~\cite{woods2014mappings,strasberg2018fermionic,iles2016energy,martinazzo2011communication,chin2010exact}, and there are direct ways of obtaining the residual spectral density without going through the intermediate steps as above.

The starting point for such analyses is the exciton spectral density {of the full set of exciton states}:
\begin{align}
\label{eq:spectral-density-definition}
J(\omega) = \sum_{\alpha\mathbf{k}} \abs{g_{\alpha\mathbf{k}}}^2\delta(\omega-\omega_\mathbf{k}){,}
\end{align}
{from which} 
we can derive the necessary quantities related to the exciton reaction coordinate. Starting with the reaction coordinate frequency, we have
\begin{align}
\Omega_{0} = \sum_{\alpha\mathbf{k}} U_{0\alpha\mathbf{k}}^*\omega_\mathbf{k} U_{0\alpha\mathbf{k}} = \frac{\int\dd{\omega} \omega J(\omega)}{\int\dd{\omega} J(\omega)},
\end{align}
which is simply the first moment of $J(\omega)$. The coupling strength can be calculated similarly,
\begin{align}
\label{eq:G0-relation-1}
G_0 = \sqrt{\sum_{\alpha\mathbf{k}}\abs*{g_{\alpha\mathbf{k}}}^2} = \sqrt{\int\dd{\omega}J(\omega)}.
\end{align}
The residual spectral density is more involved. Here we shall not prove the relation between $J$ and $J_\mathrm{res}$, but state a result from Ref.~\onlinecite{woods2014mappings},
\begin{align}
\label{eq:res-spec-dens}
J_{\rm res}(\omega) = \frac{G_0^2 J(\omega)}{\Phi^2(\omega) + \pi^2J^2(\omega)},
\end{align}
where
\begin{align}
\Phi(\omega) = \frac{1}{2}\lim_{\ell\rightarrow 0^+}\int_{a}^b \dd{\nu}J(\nu)\qty[\frac{1}{\omega-\nu-{\text{i}}\ell} + \frac{1}{\omega-\nu+{\text{i}}\ell}]
\end{align}
is {the so-called} 
\emph{reducer} and the interval $[a,b]$ is the frequency support of $J{(\omega)}$~\cite{woods2014mappings}{; in} 
the present case, the support of the spectral density is $[\omega_0,\infty)$. When calculating the reducer numerically, it is convenient to use the form~\cite{woods2014mappings}
\begin{align}
\Phi(\omega) =   J(\omega)\ln\qty[\frac{\omega-a}{b-\omega}] - \int_{a}^b\dd{\nu}\frac{J(\nu)-J(\omega)}{\nu-\omega},
\end{align}
where the upper limit of the support, $b$, is chosen sufficiently large that $J_{\rm res}$ can be considered independent thereof.
To deepen the intuitive understanding of the reaction coordinate transformation for our 2D excitonic systems, and for practical reasons, we can rewrite the creation operator of the exciton reaction coordinate by introducing the real-space exciton operators $b_\alpha^\dagger(\mathbf{r}) = S^{-1/2}\sum_\mathbf{k} e^{-i\mathbf{k}\cdot\mathbf{r}}b_{\alpha\mathbf{k}}^\dagger$, such that
\begin{align}
B_0^\dagger = \sum_\alpha\int\dd[2]{\mathbf{r}} \psi_0^\alpha(\mathbf{r}) b_\alpha^\dagger(\mathbf{r}),
\end{align}
where $\psi_0^\alpha{(\mr)}$ is the real-space exciton reaction coordinate wave function,
\begin{align}
\label{eq:psi-0}
\psi_0^\alpha(\mathbf{r}) = -\frac{\mFt_\text{c}(\mathbf{r},z_0)\cdot\mathbf{p}_\mathrm{cv}^\alpha}{\sqrt{\sum_{\alpha'}\int\dd[2]{\mathbf{r}'}\abs*{\mFt_\text{c}(\mathbf{r}',z_0)\cdot\mathbf{p}_\mathrm{cv}^{\alpha'}}^2}}.
\end{align}
From the form of $\psi_0^\alpha{(\mr)}$, it is clear that the exciton reaction coordinate inherits its spatial distribution from the vectorial projection of the {electromagnetic} field distribution {$\mFt_\text{c}(\mr)$ onto the plane of the 2D material.}

It is useful to establish the connection between the coupling strength $G_0$, and the {electromagnetic field distribution given by ${{\mFt}_\text{c}}(\mr)$}. This can be obtained by using the general exciton--field coupling coefficient, Eq.~\eqref{eq:g-k-p.A}, along with the expression for $G_0$, Eq.~\eqref{eq:G0-relation-1}, which leads directly to
\begin{align}
\hbar G_0 =\sqrt{\frac{\hbar e_0^2}{\pi\epsilon_0 m_0^2 {\omega_\text{c}} a_{\rm B}^2}\sum_{\alpha}\int\dd[2]{\mathbf{r}}\abs*{{{\mFt}_\text{c}}(\mathbf{r},z_0)\cdot\mathbf{p}_\mathrm{cv}^\alpha}^2}.
\label{Eq:G0_real_space_mode_function_integral}
\end{align}
Importantly, the coupling strength is independent of the lateral confinement length of the optical mode. This can already be seen in Eq.~\eqref{Eq:G0_real_space_mode_function_integral}, where the integral over the mode extends over the entire 2D surface. The independence of $G_0$ on the lateral confinement stems from the fact that the exciton reaction coordinate wave function, Eq.~\eqref{eq:psi-0}, is perfectly matched with the optical mode within the 2D plane. In Sec.~\ref{sec:separable-cavity-mode}, we show this analytically for the case of a mode profile that is separable in the in-plane and out-of-plane coordinates and see that the coupling strength only depends on the out-of-plane confinement length. In Appendix~\ref{sec:variation-nanorod-length}, we have also performed numerical calculations with a QNM of a gold nanorod resonator as in Fig.~\ref{fig:intro}, where the resonator length is varied. The calculations show that the contribution to $G_0$ from the integral over the normalised QNM profile in Eq.~\eqref{Eq:G0_real_space_mode_function_integral} varies only about 1\% when the length of the nanorod - and hence the lateral confinement length of the QNM - is varied between 80 nm and 95 nm.

\subsection{Exciton-exciton interactions within the reaction coordinate}
\label{sec:excit-excit-inter}
The exciton-exciton interaction, $\hat{W}$, can be rewritten in 
{the transformed basis of exciton modes $\{\hat{B}_i\}$ as}
\begin{align}
  \begin{split}
\hat{W} = \sum_{\alpha\mathbf{k}\mathbf{k}'\mathbf{q}}\sum_{ii'jj'} &\hbar W_{\mathbf{k}\mathbf{k}'\mathbf{q}} U_{i\alpha\mathbf{k}+\mathbf{q}}U_{j\alpha\mathbf{k}'-\mathbf{q}}  U_{j'\alpha\mathbf{k}'}^*  U_{i'\alpha\mathbf{k}}^* \\ &\times\hat{B}_i^\dagger \hat{B}_j^\dagger \hat{B}_{j'}\hat{B}_{i'}.
\end{split}
\end{align}
In {the} 
dynamical model derived below, we account only for the exciton-exciton interactions within the exciton reaction coordinate{. This is justified} 
by the assumption that this is the only region of the exciton Hilbert space where 
the exciton density is sufficiently large to give a significant contribution to the dynamics. Thus, keeping only the term $i,i',j,j'=0$ in the summation, we end up with the interaction term
\begin{align}
\hat{W}_0 = \hbar W_0' \hat{B}_0^\dagger \hat{B}_0^\dagger \hat{B}_0 \hat{B}_0,
\end{align}
with the effective nonlinear interaction strength
\begin{align}
W_0' = \sum_{\alpha\mathbf{k}\mathbf{k}'\mathbf{q}} W_{\mathbf{k}\mathbf{k}'\mathbf{q}}U_{0\alpha\mathbf{k}+\mathbf{q}}U_{0\alpha\mathbf{k}'-\mathbf{q}}  U_{0\alpha\mathbf{k}'}^*  U_{0\alpha\mathbf{k}}^*.
\end{align}
When the characteristic confinement length of the {electromagnetic field} 
is large compared with the exciton Bohr radius, the transformation elements $U_{0\mathbf{k}}$ decay on a momentum scale that is small compared to the momentum variation of $W_{\mathbf{k}\mathbf{k}'\mathbf{q}}$. In this regime, we can approximate $W_{\mathbf{k}\mathbf{k}'\mathbf{q}}\simeq W_{000}$. Furthermore, using the spatial reaction coordinate wave function, $\psi_0$, the effective nonlinear interaction strength can be vastly simplified as
\begin{align}
  \label{eq:exciton-exciton-interaction-reaction-coordinate}
W_0' = SW_{000}\sum_\alpha\int\dd[2]{\mathbf{r}} \abs*{\psi_0^\alpha(\mathbf{r})}^4.
\end{align}
This result is consistent with Ref.~\onlinecite{verger2006polariton}, where it is assumed that there exists an excitonic eigenmode with the same spatial wave function as the resonator mode. As we shall see explicitly later, the integral of $\abs*{\psi_0^\alpha(\mathbf{r})}^4$ is a measure of only the lateral confinement of the optical mode. Thus, the lateral optical confinement is inherited by the exciton reaction coordinate, thereby determining the interaction strength between excitons within the reaction coordinate.

\subsection{Localized and separable mode profiles}
\label{sec:separable-cavity-mode}
{In this subsection, we investigate the coupling dynamics in an idealized limit of an electromagnetic mode function that is localized and separable. We do this in order to illustrate the mechanism by which the perfect co-localization of the exciton reaction coordinate and the resonant field leads to a coupling strength that is independent of the lateral extent of the electromagnetic field.

The electromagnetic field distribution $\mFt_\text{c}(\mr)$ is localized in space, as discussed in Section~\ref{sec:resonant-electromagnetic-fields}. 
It is advantageous, therefore, to think of $\mFt_\text{c}(\mr)$ as a hypothetical localized solution to the wave equation with a purely real frequency. Even if all resonant electromagnetic modes of optical cavities and plasmonic particles have finite $Q$-values and are leaky in nature, the abstraction of perfect temporal and spatial confinement allows us to derive a number of interesting analytical results regarding the effect of the electromagnetic field distribution on the coupling strength. At positions close to the resonator, 
$\mFt_\text{c}(\mr)$ closely resembles the QNM $\mft_\text{c}(\mr)$\, since they are  related by the analytical continuation $\tlo_\text{c}\rightarrow\omega_\text{c}$. Therefore, 
we expect the general findings to apply qualitatively also to electromagnetic resonators with finite $Q$-values.

With this motivation, we
consider now a localized mode function $\mathbf{\tilde{F}}_{\rm c}(\mr)$ in an environment of constant and real permittivity $\epsilon$, and we assume that the field is}
%
separable in the lateral and perpendicular coordinates, such that
\begin{align}
\mathbf{\tilde{F}}(x,y,z) = \mathbf{n} \tilde{F}_z(z)\tilde{F}_{\|}(x,y),
\label{Eq:separabel_mode_functcion}
\end{align}
where $\mathbf{n}$ is the unit polarisation vector of the mode. Since $\mathbf{\tilde{F}}(\mr)$ is localized in space, it
obeys the normalization requirement
\begin{align}
  \label{eq:normalisation-separable-mode}
\epsilon_{\rm eff}\int \dd{z}\abs*{\tilde{F}_z(z)}^2\int\dd[2]{\mathbf{r}} \abs*{\tilde{F}_{\|}(\mathbf{r})}^2=1\rev{,}
\end{align}
\rev{in which} 
$\epsilon_{\rm eff}$ is an effective dielectric constant \rev{to account} 
for the dielectric response of the surrounding material. While $\epsilon(\mr)$ can \rev{be considered} 
constant within the 2D semiconductor sheet, it generally varies across the surrounding structure. \rev{In order to focus on the} 
influence of the confinement length scales, \rev{however, we approximate the combined effect by use of an} 
effective dielectric constant $\epsilon_{\rm eff}$. We note that this assumption is not generally a requirement for using the theory, and for the numerically calculated QNM that we describe in Sec.~\ref{sec:benchm-with-semicl} and App.~\ref{sec:variation-nanorod-length}, we do not make any such simplifying assumptions about the dielectric environment.








\rev{Using the separable 
mode function in Eq.~(\ref{Eq:separabel_mode_functcion}) and} the normalization reguirement in Eq.~\eqref{eq:normalisation-separable-mode}, we can demonstrate that $G_0$ is entirely independent of the lateral mode distribution by rewriting it as
%
\begin{align}
\label{eq:separable-mode-G0}
  \hbar G_0 = \sqrt{\sum_\alpha\frac{\hbar e_0^2\abs*{\mathbf{n}\cdot\mathbf{p}_\mathrm{cv}^\alpha}^2}{\pi\epsilon_0  m_0^2 \omega_\text{c} a_{\rm B}^2 L_z}},
\end{align}
where
\begin{align}
L_z = \frac{\epsilon_{\rm eff}\int \dd{z}\abs*{{\tilde{F}}_z(z)}^2}{\abs*{{\tilde{F}}_{z}(z_0)}^2}
\end{align}
is the out-of-plane confinement length. It follows from Eq.~(\ref{eq:separable-mode-G0}) that the coupling strength is independent of the lateral mode distribution, and that it scales with the out-of-plane confinement length as $G_0\propto L_z^{-1/2}$. For ease of notation, we shall use an implicit summation over the valley index as $\abs*{\mathbf{n}\cdot\mathbf{p}_{\rm cv}} = (\sum_\alpha \abs*{\mathbf{n}\cdot\mathbf{p}_{\rm cv}^\alpha}^2)^{1/2}$ when stating the polarisation overlap used in specific calculations.

The reaction coordinate exciton--exciton interaction strength {$W_0'$} can be written using Eq.~\eqref{eq:exciton-exciton-interaction-reaction-coordinate} as
\begin{align}
  \label{eq:separable-mode-W0}
  \hbar W_0'  &= \hbar S W_{000}\eta_{\mathbf{n}}\frac{\int\dd[2]{\mathbf{r}} \abs*{F_{\|}(\mathbf{r})}^4}{\qty(\int\dd[2]{\mathbf{r}} \abs*{F_{\|}(\mathbf{r})}^2)^2},
\end{align}
where
\begin{align}
  \eta_{\mathbf{n}} = \frac{\sum_\alpha\abs*{\mathbf{n}\cdot\mathbf{p}_{\rm cv}^\alpha}^4}{\qty(\sum_\alpha\abs*{\mathbf{n}\cdot\mathbf{p}_{\rm cv}^\alpha}^2)^2}
\end{align}
is a polarisation-dependent prefactor. This prefactor takes values between 1/2 and 1. The maximal value, 1, is obtained when $\mathbf{n}$ is orthogonal to one of the momentum matrix elements, i.e. $\mathbf{n}\cdot\mathbf{p}_{\rm cv}^\alpha=0$. Conversely, the minimal value, 1/2, is obtained when $\abs*{\mathbf{n}\cdot\mathbf{p}_{\rm cv}^\alpha}$ is equal for the two polarisations.

\subsubsection{Gaussian lateral field distribution}
\label{sec:gauss-later-field}

\begin{figure}
  \centering
  \includegraphics[width=\columnwidth]{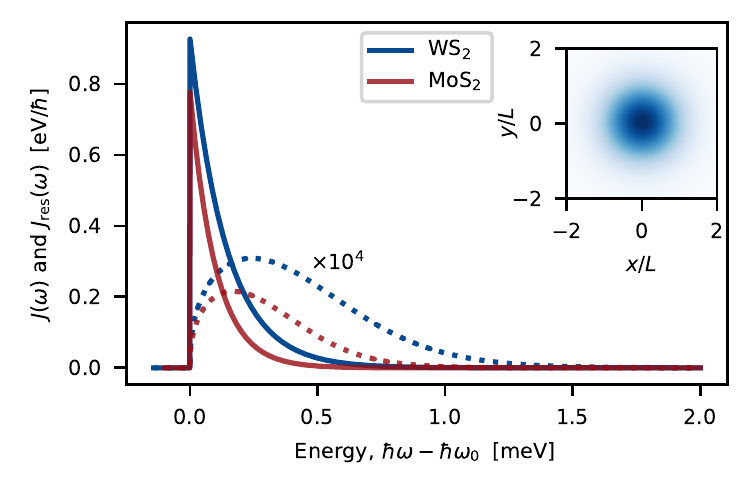}
  \caption{Excitonic spectral density (solid lines) and corres\-ponding residual spectral densities (dotted lines, both scaled up by a factor of $10^{4}$) for a Gaussian lateral field distribution 
with $L=20\mathrm{\:nm},\; L_z=200\mathrm{\:nm}$ and $\abs{\mathbf{p}_{\rm cv}\cdot\mathbf{n}}/p_{\rm cv} = 0.3$, coupled to monolayer $\mathrm{WS_2}$ (dark blue) and $\mathrm{MoS_2}$ (light blue), respectively. The inset shows the Gaussian lateral field distribution $\tilde{F}_{\|}(x,y)$.}
  \label{fig:spectral-density-gaussian}
\end{figure}

We now proceed by assuming that the lateral field distribution defined by the electromagnetic resonator 
is Gaussian with a confinement length scale $L$,
\begin{align}
\tilde{F}_{\|}(x,y) = \frac{e^{-(x^2+y^2)/(2{L}^2)}}{L\sqrt{\pi}}.
\label{Eq:gaussian_field}
\end{align}
We note that $\int\dd[2]{\mathbf{r}} \abs*{\tilde{F}_{\|}(\mathbf{r})}^2 = 1$, so that 
the normalisation requirement is $\epsilon_{\rm eff}\int \dd{z} \abs*{\tilde{F}_z(z)}^2=1$, and the
exciton coupling strengths {in Eq.~(\ref{eq:g-k-p.A}) can be written as} 
\begin{align}
g_{\alpha\mathbf{k}} = -\sqrt{\frac{4 e_0^2\abs{\mathbf{n}\cdot\mathbf{p}_\mathrm{cv}^\alpha}^2 {L}^2}{\hbar\epsilon_0m_0^2{\omega_\text{c}}a_{\rm B}^2S}  }\tilde{F}_z(z_0)e^{-\frac{1}{2}(k{L})^2}.
\end{align}
{This relatively simple expression for the coupling strength allows us to evaluate the expression for the spectral density in Eq.~\eqref{eq:spectral-density-definition} by 
writing the summation over $\mathbf{k}$ 
as an integral, $\sum_\mathbf{k}\rightarrow \frac{S}{(2\pi)^2}\int\dd[2]{\mathbf{k}}$. In this way, we find that we can write} 
the spectral density {compactly in terms of the reaction coordinate coupling strength in Eq.~(\ref{eq:separable-mode-G0})} as
\begin{align}
  \label{eq:sd-gaussian}
  J(\omega) = (G_0^2/\xi)\Theta(\omega-\omega_0) e^{-(\omega-\omega_0)/\xi},
\end{align}
%
%
where $\xi = \hbar/(2M{L}^2)$ is a cutoff frequency and $\Theta$ is the Heaviside function.
{Using this analytical expression for the} spectral density, it is also possible to evaluate the residual spectral density in Eq.~(\ref{eq:res-spec-dens}) analytically,
\begin{align}
  \label{eq:res-sd-gaussian}
J_\mathrm{res}(\omega) = \frac{\xi\Theta(\omega-\omega_0)e^{(\omega-\omega_0)/\xi}}{\mathrm{Ei}^2[(\omega-\omega_0)/\xi] + \pi^2},
\end{align}
where $\mathrm{Ei}(x)=\int_{-\infty}^x\dd{z}\exp(z)/z$ is the exponential integral function.

Fig.~\ref{fig:spectral-density-gaussian} shows an example of the spectral density and residual spectral density generated by {the Gaussian field distribution in Eq.~(\ref{Eq:gaussian_field})} 
coupled to $\mathrm{WS_2}$ and $\mathrm{MoS_2}$, respectively. As can be seen from {Eq.~\eqref{eq:sd-gaussian},} 
the magnitude of the spectral density 
{scales} as the ratio $G_0^2/\xi$. Here, $G_0$ depends on the material-specific parameters ${\mathbf{p}}_{\rm cv}^\alpha$ {and} $a_{\rm B}$ {along with} 
the 
{out-of-plane} confinement length scale $L_z${. The} 
cutoff frequency {$\xi$} depends on the material-specific total exciton mass {$M$} as well as the in-plane confinement length scale of the {resonator, ${L}$}. The magnitude of the residual spectral density, on the other hand, scales as the cutoff frequency $\xi$. Thus, the cutoff frequency determines not only the relevant frequency scale of the exciton spectral density, but also the relative strength of the interactions with the residual exciton environment.

The reaction coordinate exciton--exciton interaction strength $W_0'$ can be obtained from Eq.~\eqref{eq:separable-mode-W0} as
\begin{align}
  \hbar W_0' 
  &= \frac{\hbar S W_{000}\eta_{\mathbf{n}}}{2\pi {L}^2},
\end{align}
As anticipated in Sec.~\ref{sec:introduction}, this nonlinear interaction strength scales inversely with the confinement area, ${L}^2$, reflecting that the co-localisation of multiple excitons (and thereby their interaction) is fully determined by the electromagnetic field distribution.
For the calculations in this article that involve nonlinear interactions with Gaussian electromagnetic modes (Sec.~\ref{sec:pulse-driv-nonl}), we take the polarisation prefactor to be unity, corresponding to a circularly polarised optical mode.

\section{Time evolution}
\label{sec:time-evolution}
Based on the general framework in Sec.~\ref{sec:general-framework} and the reformulation in terms of the reaction coordinate in Sec.~\ref{sec:reaction-coordinate}, we are now in a position to calculate the time evolution of the excitations in the coupled system comprising the electromagnetic resonator and the 2D material.

In this section, we {present} three strategies {of increasing complexity for calculating the system dynamics.} 
The first method{, which} is exact {when} excitonic broadening effects are ignored{, is based on direct time evolution of the single-excitation product states formed by the Fock states of a single electromagnetic excitation and the continuum of single-exciton states with given momentum.
The second method is based on a Markovian treatment of the residual exciton modes, and a master equation for the reduced density operator of the {resonant electromagnetic field and the exciton} reaction coordinate is derived. {In this formulation, one can account for} 
excitonic decay and dephasing within the reaction coordinate{,} 
and external driving of the system can be included as well. The Markovian master equation {is} 
benchmarked against the exact treatment when excitonic broadening effects and external driving are ignored, thereby providing a reference calculation to assess the Markov approximation for the interaction with the residual exciton modes. The third method is based on an iterative extension of the reaction coordinate mapping, allowing {one} to represent the residual exciton modes by a one-dimensional chain of bosonic modes. This{, in turn, enables the model to} 
account for 
non-Markovian features of the residual environment, although our implementation can only evolve the system up to a finite time. Furthermore, using this approach, it is also straightforward to include external driving with a time-dependent amplitude, as is the case when the system is excited by a laser pulse.

\subsection{Exact evolution in the single-excitation sector}
\label{sec:exact-evol-single}
{As a relatively simple starting point, and to} 
establish a benchmark for evaluating the precision of {more complicated} approaches for {time-evolution calculations, we consider in this section the Fock state representation of a simplified model system. This approach is motivated by the fact} 
that the dynamics can be solved exactly, when the coupled system is restricted to the one-excitation sector. In practice, we {do} this by initialising the resonant electromagnetic field in a single-photon Fock state, such that nonlinearities can be neglected. We can then expand the combined state as
\begin{align}
\ket{\Psi(t)} = \phi_{\text{c}}(t)\ket{1;\{0\}} + \sum_{\alpha\mathbf{k}} \phi_{\alpha\mathbf{k}}(t)\ket{0;1_{\alpha\mathbf{k}}},
\end{align}
where $\ket{1;\{0\}}$ denotes the state with a single electromagnetic excitation and zero excitons and $\ket{0;1_{\alpha\mathbf{k}}}$ denotes the state with no electromagnetic excitations combined with a single exciton with momentum $\mathbf{k}$. The amplitude $\phi_{\rm c}$ then obeys the equation of motion~\cite{vats2002theory}
\begin{align}
\label{eq:exact-time-evolution}
  \dv{\phi_{\text{c}}(t)}{t} = -\int_0^t\dd{t'} K(t-t')\phi_{\text{c}}(t') - \gamma_{\text{c}}\phi_\text{c}(t),
\end{align}
where
\begin{align}
K(\tau) = \Theta(\tau)\int\dd{\omega} J(\omega) e^{-\mathrm{i}(\omega-{\omega_\text{c}})\tau}.
\end{align}
{Compared to Ref.~\onlinecite{vats2002theory}, the roles of the electromagnetic and electronic degrees of freedom are interchanged, so that the excitons act as a continuum with which the single electromagnetic field of the resonator interacts. Since $J(\omega)$ does not suffer from an ultraviolet divergence, there is no need to introduce a cut-off in order to evaluate the integral. Although this method is in principle exact, 
it is limited to single-excitation problems and cannot be used to model problems with 
external driving or dephasing effects. For this reason, we use the method mostly for reference calculations for the master equation formulations to be described below.} 

\subsection{Secular Markovian master equation}
\label{sec:markovian-master-equation}
{In this section we} 
derive a secular Markovian master equation for the reduced density operator {$\hat{\rho}$} of the system comprising the {resonant electromagnetic field} 
and the exciton reaction coordinate, 
{where} the residual exciton modes are traced out.
We {subsequently} 
benchmark this master equation {approach} against the exact method {from section~\ref{sec:exact-evol-single}}. We do this in the limit where dephasing and driving effects are turned off in order to assess the accuracy of the master equation and study the conditions for describing the interactions with the residual exciton modes within the Markov approximation.


In the Hamiltonian governing the evolution of the resonant electromagnetic field, we include external laser driving with frequency $\omega_{\rm d}$ and constant amplitude $F$ in addition to the free evolution described in Section~\ref{sec:resonant-electromagnetic-fields}, such that the part of the Hamiltonian governing the field evolution is $\hat{H}_{\rm c} = \hbar\omega_{\rm c}\hat{a}_{\rm c}^\dagger\hat{a}_{\rm c}+ \hbar F(e^{\mathrm{i}\omega_{\rm d}t}\hat{a}_{\rm c} + e^{-\mathrm{i}\omega_{\rm d}t}\hat{a}_{\rm c}^\dagger)$.

Next, we divide the total Hamiltonian into contributions describing the exciton reaction coordinate and the resonant mode, being the system (S), the residual exciton modes, being the reservoir (R), and their mutual coupling (SR), as $\hat{H} = \hat{H}_{\text{S}} + \hat{H}_{\text{R}} + \hat{H}_{\text{SR}}$. 
Putting it all together, we have
\begin{align}
\label{eq:H-S-R-SR}
\begin{split}
\hat{H}_{\text{S}} &= \hbar ({\omega_\text{c}} -\omega_{\text{d}}) \hat{a}_{\rm c}^\dagger \hat{a}_{\rm c}  + \hbar(\Omega_0 - \omega_{\text{d}}) \hat{B}_0^\dagger \hat{B}_0 + \hat{W}_0 \\ &+ \hbar G_0(\hat{B}_0^\dagger \hat{a}_{\rm c} + \hat{B}_0 \hat{a}_{\rm c}^\dagger) + \hbar F (\hat{a}_{\rm c}+\hat{a}_{\rm c}^\dagger),\\
\hat{H}_{\text{R}} &= \sum_{i>0} \hbar (\tilde{\Omega}_{i}-\omega_{\text{d}}) \hat{\tilde{B}}_i^\dagger \hat{\tilde{B}}_i \\
\hat{H}_{\text{SR}} &= \sum_{i>0} \hbar\tilde{\lambda}_i \hat{B}_0 \hat{\tilde{B}}_i^\dagger + \hbar\tilde{\lambda}_i^* \hat{B}_0^\dagger \hat{\tilde{B}}_i,
\end{split}
\end{align}
in which we have expressed the Hamiltonian in 
a reference frame rotating with the driving frequency $\omega_{\text{d}}$. The standard Markovian master equation obtained by tracing out the environment is given by~\cite{breuer2002theory}
\begin{align}
\label{eq:markovian-me-1}
\dv{\hat{\rho}}{t} = -\frac{{\text{i}}}{\hbar}[\hat{H}_{\text{S}},{\hat{\rho}}] + {2}\gamma_{\text{c}}\mathcal{D}(\hat{a}_{\rm c},\hat{\rho}) +  \mathcal{K}[{\hat{\rho}}],
\end{align}
{in which the Lindblad dissipator $\mathcal{D}(\hat{x},\hat{\rho})$ is specified in Eq.~(\ref{eq:lindblad-dissipator}),} 
and
\begin{align}
\label{Eq:exciton_dissipator_starting_point}
\mathcal{K}[{\hat{\rho}}] = -\frac{1}{\hbar^2}\int_0^\infty\dd{\tau} \tr_{\text{R}}[\hat{H}_{\text{SR}},[\hat{H}_{\text{SR}}(-\tau),{\hat{\rho}}\otimes\hat{\rho}^0_{\text{R}}]],
\end{align}
where 
$\hat{H}_{\text{SR}}(-\tau)=e^{-{\text{i}}(\hat{H}_{\text{S}}+\hat{H}_{\text{R}})\tau/\hbar}\hat{H}_{\text{SR}}e^{+{\text{i}}(\hat{H}_{\text{S}}+\hat{H}_{\text{R}})\tau/\hbar}$ {is} the interaction-picture time evolution of $\hat{H}_{\text{SR}}$ and $\hat{\rho}^0_{\text{R}}$ {is} the initial state of the residual excitonic environment, which we take to be 
the exciton vacuum. 
As discussed in Appendix~\ref{sec:exciton-dissipator}, we can use the secular approximation to write the exciton dissipator $\mathcal{K}[{\hat{\rho}}]$ in a simpler form as
  \begin{align}
\label{eq:secular-markovian-dissipator-full}
\begin{split}
  \mathcal{K}[\hat{\rho}] = -\sum_\omega &\Big\{\Gamma_\mathrm{res}(\omega)\mathcal{D}[\hat{B}_0(\omega),\hat{\rho}]\\ & - \pk{\text{i}}\Delta_\mathrm{res}(\omega)[ (\hat{B}_0(\omega))^\dagger \hat{B}_0(\omega),\hat{\rho}]\Big\},
\end{split}
\end{align}
where the sum is over all system eigenfrequency differences $\omega$, and 
the exciton dissipation rate $\Gamma_\text{res}(\omega)$ can be written in terms of the residual spectral density as
\begin{align}
\label{eq:secular-markovian-decay-rate}
  \Gamma_\mathrm{res} (\omega) = 2\pi J_\mathrm{res} (\omega+\omega_{\text{d}}).
\end{align}
The operators $\hat{B}_0(\omega)$ and $\hat{B}_0^\dagger(\omega)$ are eigenstate-projected exciton operators, which are described in detail in App.~\ref{sec:exciton-dissipator}, where the term $\Delta_{\rm res}(\omega)$ is also defined.
In the weak-driving limit, $F \ll G_0, \gamma_{\text{c}}$, only the low-energy states of the system are populated, and $G_0$ dominates the structure of the coupled system. In this limit, 
we can calculate the eigenstates of $\hat{H}_{\text{S}}$ in the basis $\{\ket{0,0},\ket{1,0},\ket{0,1}\}$, where $\ket{n_{\text{c}},n_{\text{x}}}$ denotes a Fock state with $n_{\text{c}}$ {electromagnetic energy quanta} 
in the {resonant field mode} 
and $n_{\text{x}}$ excitons in the reaction coordinate mode. We find that there are only two non-zero contributions to the summation over $\omega$, corresponding to the upper and lower polariton modes, with frequencies
\begin{align}
\begin{split}
\bar{\omega}_+ &= \frac{1}{2}\qty({\omega_\text{c}} + \Omega_0 - 2\omega_{\text{d}} + \eta),\\
\bar{\omega}_- &= \frac{1}{2}\qty({\omega_\text{c}} + \Omega_0 - 2\omega_{\text{d}} - \eta). \\\: \\
\end{split}
\end{align}
The corresponding eigenstate-projected operators are the annihilation operators of the upper and lower polaritons,
\begin{align}
  \begin{split}
\hat{B}_0(\bar{\omega}_+) &= \frac{2G_0(-\delta_{\text{cx}} + \eta)}{4G_0^2 + (-\delta_{\text{cx}} + \eta)^2} \hat{a}_{\rm c}
+ \frac{(-\delta_{\text{cx}} + \eta)^2}{4G_0^2 + (-\delta_{\text{cx}} + \eta)^2}\hat{B}_0 \\ \: \\
\hat{B}_0(\bar{\omega}_-) &= -\frac{2G_0(\delta_{\text{cx}} + \eta)}{4G_0^2 + (\delta_{\text{cx}} + \eta)^2} \hat{a}_{\rm c}
+ \frac{(\delta_{\text{cx}} + \eta)^2}{4G_0^2 + (\delta_{\text{cx}} + \eta)^2}\hat{B}_0
\end{split}
\end{align}
where $\delta_{\text{cx}}={\omega_\text{c}}-\Omega_0$ is the exciton--resonator detuning and $\eta = \sqrt{4G_0^2 + \delta^2_{\text{cx}}}$ is the polariton splitting. The bars over $\bar{\omega}_{\pm}$ signify that the polariton frequencies are given in the rotating frame. They are related to the corresponding lab frame frequencies $\omega_{\pm}$ as $\omega_{\pm}=\bar{\omega}_{\pm}+\omega_{\rm d}$. In the case where the lower-polariton energy $\omega_-$ is below the exciton gap $\omega_0$, (at resonance, this amounts to $G_0>(\Omega_0-\omega_0)$) the contribution to the dissipator from $\hat{B}_0(\bar{\omega}_-)$ vanishes, and we can write the master equation in the simplified form
\begin{align}
\label{eq:markovian-me-simple}
\dv{\hat{\rho}}{t} = -\frac{{\text{i}}}{\hbar}[\hat{H}_{\text{S}},\hat{\rho}] + {2}\gamma_{\text{c}}\mathcal{D}(\hat{a}_{\rm c},\hat{\rho}) + \Gamma_\mathrm{res}(\bar{\omega}_+)\mathcal{D}(\hat{B}_0',\hat{\rho}),
\end{align}
where $\hat{B}_0':=\hat{B}_0(\bar{\omega}_+)$.

These derivations put us in a position to understand the impact of the different energy scales in the system on a deeper level. In Fig.~\ref{fig:residual-decoupling-schematic}, the residual spectral density is shown along with {an} 
indication of the reaction coordinate frequency {$\Omega_0$}. The upper and lower polariton frequencies {$\omega_\pm$} are also indicated for
the resonant case ${\omega_\text{c}}=\Omega_0$, where $\omega_\pm = \Omega_0\pm G_0$. From Eq.~\eqref{eq:secular-markovian-decay-rate}, we know that the dissipation rates from the polaritons into the residual environment are given by $2\pi J_{\rm res}(\omega_\pm)$. Thus, when the coupling strength is increased far beyond the cutoff frequency {$\xi$}, the polaritonic peaks are pushed away from the peak of $J_{\rm res}{(\omega)},$ and the effective interaction strength with the residual excitons is {consequently} reduced. Correspondingly, if the cutoff frequency is decreased, for example by increasing the lateral confinement length {$L$}, the same reduction in effective interaction strength is observed. In the following section, we shall see that this phenomenon of effectively decoupling the residual excitons when $G_0\gg\xi$ leads to a large parameter regime where the residual environment can be safely ignored.

\begin{figure}
  \centering
  \includegraphics[width=\columnwidth]{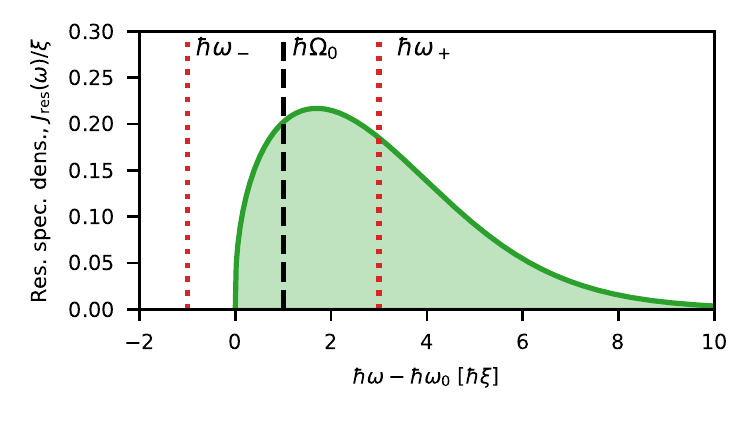}
  \caption{Residual spectral density (green shaded area) for a Gaussian optical mode shown along with the exciton reaction coordinate frequency, $\Omega_0$, and the polariton frequencies, $\omega_\pm$ (orange dotted lines), for the resonant case, ${\omega_\text{c}}=\Omega_0$ and for $G_0=2\xi$. The indicated frequencies are shown relative to the exciton gap {$\omega_0$}. In the Markovian master equation, the dissipation rates from the polaritons into the residual exciton environment are proportional to $J_{\rm res}(\omega_\pm)$. Thereby, a large coupling strength, $G_0$, relative to the cutoff frequency, $\xi$, leads to a weaker interaction with the residual environment, as the polariton peaks are displaced further away from the peak of $J_{\rm res}$.}
  \label{fig:residual-decoupling-schematic}
\end{figure}

\subsubsection{Benchmarking}
\label{sec:markov-benchmarking}

In Fig.~\ref{fig:Markovian-benchmark}a we show the temporal evolution of the resonator population for three different lateral confinement length scales. The initial state is a single excitation in the resonator. We compare the time evolution calculated by the exact equation of motion of the resonant electromagnetic field {in} Eq.~\eqref{eq:exact-time-evolution} (black solid lines) with the secular Markovian master equation with all terms included {in Eq.~(\ref{eq:secular-markovian-dissipator-full})} 
(green dots), as well as the simplified Markovian master equation {in} 
Eq.~\eqref{eq:markovian-me-simple} (dashed orange lines). For reference, we have also shown the time evolution generated when the residual excitons are entirely ignored (red dotted lines). In general, the dynamics show Rabi oscillations due to the interaction between the exciton reaction coordinate and the {resonator field.} 
These oscillations are damped due to Markovian losses 
{of the resonator} 
and interactions with the residual excitons. As discussed earlier, the effect of the residual excitons becomes more pronounced as the lateral size decreases. In Fig.~\ref{fig:Markovian-benchmark}b, the relative error{s} of the three approximate approaches {are} 
shown. The error{s are 
calculated as
\begin{align}
\revv{\mathcal{E}_{\rm rel} =} \frac{\int \dd{t} [\ev{\hat{a}_{\rm c}^\dagger(t) \hat{a}_{\rm c}(t)} - \abs{\phi_{\rm c}(t)}^2]^2}{\int \dd{t} \abs{\phi_{\rm c}(t)}^4},
\label{Eq:rel_error_calc}
\end{align}
where} $\ev{\hat{a}_{\rm c}^\dagger(t) \hat{a}_{\rm c}(t)}$ is the population of the electromagnetic resonator obtained with one of the master equations and $\phi_{\rm c}(t)$ is obtained from the exact time evolution. Furthermore, the Markovian decay rate into the residual exciton modes {$\Gamma_\mathrm{res}$} is shown (blue solid line, right $y$-axis). The dependence of the error{s} on the lateral length scales confirms that the influence of the residual excitons is 
{more pronounced} at small length scales. For the chosen parameters, we find that the residual excitons can be ignored when ${L}\gtrsim 4\mathrm{\: nm}$. Furthermore, we see that the Markovian master equation provides a useful description of the interactions with the residual environment, which improves the accuracy of the calculated time evolution. We also conclude that the error of the simplified Markovian master equation, Eq.~\eqref{eq:markovian-me-simple}, is comparable with the full Markovian dissipator, Eq.~\eqref{eq:secular-markovian-dissipator-full} with all secular terms included. In a significant portion of the parameter regime, the simplified master equation even performs {slightly} better.
\begin{figure}
  \centering
  \includegraphics[width=\columnwidth]{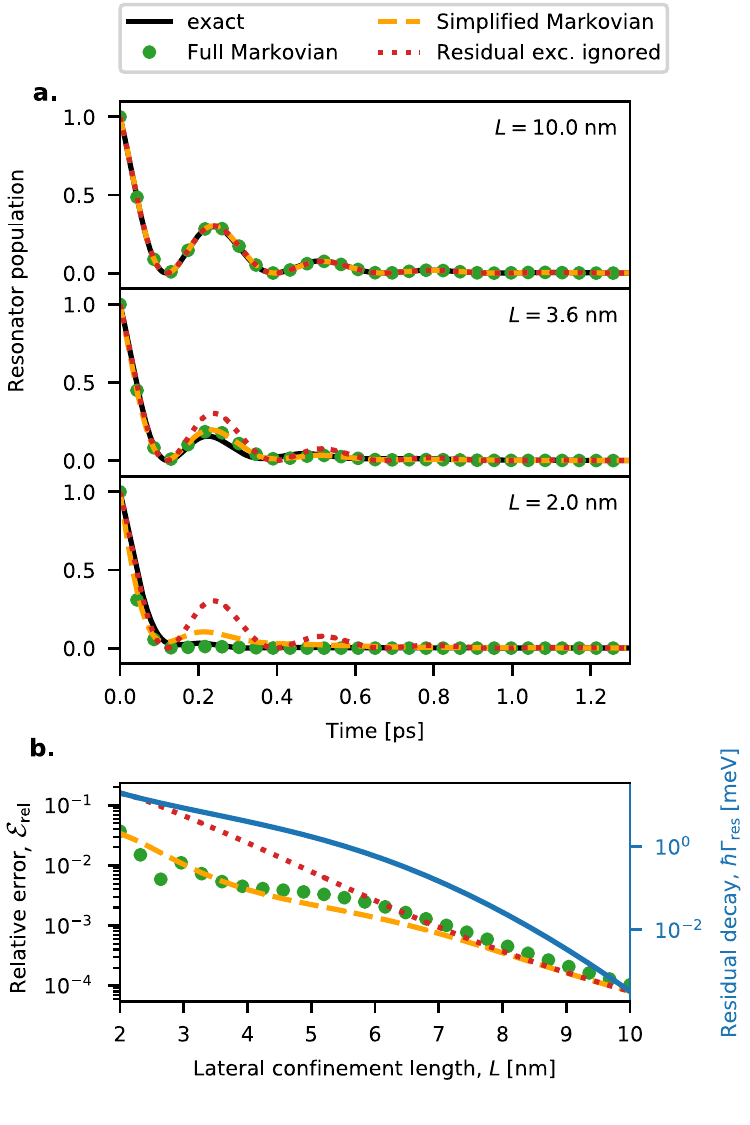}
  \caption{{\bf a.} Exact transient dynamics of Gaussian resonator mode coupled resonantly to $\mathrm{WS_2}$ (black solid lines) compared to the solution of the simplified Markovian master equation, Eq.~\eqref{eq:markovian-me-simple} (orange dashed lines) and the Markovian master equation with all secular terms included, Eq.~\eqref{eq:secular-markovian-dissipator-full} (green dots). For reference, the solution of the master equation without taking the residual exciton environment into account is also shown (dotted red lines). Parameters: $L_z=200 \mathrm{\; nm}, 2\hbar\gamma_c=6.6\mathrm{\; meV}, \; \abs{\mathbf{p}_{\rm cv}\cdot\mathbf{n}}/p_{\rm cv}=0.2$, corresponding to $\hbar G_0=7.7\mathrm{\; meV}$. {\bf b.} The relative error (left axis) of the three approximate approaches as a function of lateral confinement length, $L$, with line styles matching those in panel a. The blue solid line (right axis) shows the Markovian decay rate from the upper polariton into the residual environment.}
  \label{fig:Markovian-benchmark}
\end{figure}

\subsubsection{Inclusion of excitonic broadening effects}
\label{sec:incl-excit-broad}
\begin{table*}
  \centering
  \begin{tabular}{l|c|c|c|c}
    Material & $\gamma_0^\mathrm{nr} \;[\mathrm{meV}]$ & $c_1 \;[\mathrm{\mu eV/K}]$ & $c_2 \;[\mathrm{meV}]$ & $ \Omega\; [\mathrm{meV}]$
\\ \hline
$\mathrm{WSe_2}$ & 3.6 & 56 & 9.4 & 15 \\
$\mathrm{WS_2}$ & 2.1 & 28 & 6.5 & 20 \\
$\mathrm{MoS_2}$ & 0 & 91 & 8.4 \text{(decay)}, 7.2 \text{(dephasing)} & 30
  \end{tabular}
  \caption{Parameters from Ref.~\onlinecite{selig2016excitonic} for calculation of temperature-dependent exciton decay and dephasing rates. The parameters are the temperature-independent contribution to the non-radiative decay from phonon interactions, $\gamma_0^{\rm nr}$, the contribution to dephasing from intravalley phonon scattering, $c_1$, the contribution to decay (and for $\mathrm{MoS_2}$ also dephasing) due to intervalley phonon scattering, $c_2$, and the typical phonon energy for intervalley processes, $\Omega$. }
  \label{tab:exciton-broadening-parameters}
\end{table*}

Because of interactions with lattice phonons, the excitons experience population decay and dephasing, which result in a temperature-dependent broadening of the exciton line~\cite{dey2016optical,selig2016excitonic}. In Ref.~\onlinecite{selig2016excitonic}, it was found that the dominant mechanisms generating excitonic decay are radiative recombination (which is temperature-independent) and scattering with phonons at the $\Lambda$ point. In addition, an important contribution to the exciton linewidth was found from intra-valley scattering with phonons at the $\Gamma$ point. Specifically, the total exciton linewidth of various monolayer transition-metal dichalcogenides was found to be well described by the temperature-dependent expression $\Gamma_{\text{x}} = \gamma_0 + c_1T + c_2/[\exp\{\Omega/k_{\rm B}T\}-1]$, where $c_1,\;c_2$ and $\Omega$ are material-dependent coefficients. For $\mathrm{WS_2}$ and $\mathrm{WSe_2}$, the temperature independent term $\gamma_0$ contains contributions from radiative decay and spontaneous emission of $\Lambda$-phonons.
The linear coefficient, $c_1$, stems from intra-valley scattering with thermally excited phonons at the $\Gamma$ point, and the last term accounts for interactions with thermally excited phonons at the $\Lambda$ point. The
first and last terms describe processes that lead to a decay of excitons. 
Whereas radiative decay is accounted for through the interaction with the electromagnetic field, $\hat{H}_{\rm I}$, we include the possibility of scattering into dark exciton states 
as a decay term in {the} master equation {of the form} $2\gamma_{\text{x}}\mathcal{D}(\hat{B}_0,\hat{\rho})$, where $\gamma_{\text{x}} = \gamma_{0}^\mathrm{nr} +
c_2/[\exp\{\Omega/k_{\rm B}T\}-1]$ and $\gamma_{0}^\mathrm{nr}$ is the non-radiative contribution to $\gamma_0$. Intra-valley scattering with $\Gamma$-phonons, on the other hand, does not lead to a population decay but rather to a dephasing, similar to virtual phonon transitions to higher-lying excited states seen in systems with localised exciton states~\cite{muljarov2004dephasing,reigue2017probing,tighineanu2018phonon}. This process can be included in {the} master equation by a dephasing term, $2\gamma'_{\text{x}}\mathcal{D}(\hat{B}_0^\dagger \hat{B}_0,\hat{\rho})$, where $\gamma'_{\text{x}}=c_1T$. Thus, although the total exciton linewidth, $\Gamma_{\text{x}}=\gamma_{\text{x}}+\gamma'_{\text{x}}$, depends only on the sum of the two
contributions, it is important to note that there is a difference in nature between population decay and dephasing processes. In Ref.~\onlinecite{selig2016excitonic}, monolayer ${\rm MoS_2}$ {was} 
also studied{, and it was found that} 
the coefficient $c_2$ contains contributions both from intra-valley scattering, i.e. dephasing, and from interactions with $\Lambda$ phonons, i.e. decay into dark exciton states. The parameters from Ref.~\onlinecite{selig2016excitonic} are presented in Table~\ref{tab:exciton-broadening-parameters}.

We note that recent studies indicate strain as a possible way of energetically shifting the direct 
$K$-valley exciton 
below the indirect, momentum-dark $K-\Lambda$ exciton in monolayer $\mathrm{WSe_2}$~\cite{hsu2017evidence}. Such an energetic cross-over would arguably lead to significant reduction of phonon-induced decay of the bright exciton, even with a vanishing decay at zero temperature, as is the case for $\mathrm{Mo}$-based transition-metal dichalcogenides, where the $K-\Lambda$ exciton is below the direct $K$-exciton in the absence of strain.

\subsection{Non-Markovian treatment of residual excitons using chain mapping}
\label{sec:chain-mapping}
In some situations, e.g. for small lateral confinement scales, it may be necessary to account for non-Markovian effects in the interaction with the residual excitonic environment. This can be done by extracting additional reaction coordinates from the residual environment, thereby extending the system Hilbert space to include the most important environmental degrees of freedom. Since the exciton reaction coordinate Hamiltonian {in} Eq.~\eqref{eq:H-exc-RC-2} is structurally equivalent to the original Hamiltonian, we can iterate the procedure of extracting reaction coordinates, thereby generating a one-dimensional chain of coupled modes{, as illustrated in} 
Fig.~\ref{fig:chain-mapping-illustration}. 
This strategy has been formally studied in the literature~\cite{chin2010exact,woods2014mappings}; here, we derive it iteratively, starting from the reaction coordinate transformation. First, we re-write Eq.~\eqref{eq:H-exc-RC-2} by adding the superscript {`$(0)$'} to the residual modes and quantities related to them,
\begin{align}
  \begin{split}
\hat{H}_{\rm x,0} = &\hbar\Omega_0 \hat{B}_0^\dagger \hat{B}_0 + \sum_{i>0} \hbar\tilde{\Omega}_i^{(0)} \hat{\tilde{B}}_i^{(0)\dagger} \hat{\tilde{B}}_i^{(0)} \\ &+ \sum_{i>0}\hbar\tilde{\lambda}_i^{(0)}\hat{B}_0\hat{\tilde{B}}_i^{(0)\dagger} + \hbar\tilde{\lambda}_i^{(0)*}\hat{B}_0^\dagger \hat{\tilde{B}}_i^{(0)}.
\end{split}
\end{align}
{The} 
superscript {`$(0)$'} indicates that the residual environment is coupled to the reaction coordinate $\hat{B}_0$. Correspondingly, we label the spectral density of the $\hat{B}_i^{(0)}$-modes by $J_\mathrm{res}^{(0)} = \sum_{i>0} \abs*{\tilde{\lambda}_i^{(0)}}^2\delta(\omega-\tilde{\Omega}^{(0)}_i)$. We can now define a new reaction coordinate
\begin{align}
 \hat{B}_1 = \Big[\sum_{i>0}\abs*{\tilde{\lambda}_i^{(0)}}^2 \Big]^{-1/2}\sum_{i>0} \tilde{\lambda}_i^{(0)} \hat{\tilde{B}}_i^{(0)}.
\end{align}
By repeating the procedure of defining a new residual reservoir and re-diagonalising it, as in Sec.~\ref{sec:reaction-coordinate}, we obtain
\begin{align}
  \begin{split}
    &\hat{H}_{\rm x,0} = \hbar\Omega_0 \hat{B}_0^\dagger \hat{B}_0 + \hbar\Omega_1 \hat{B}_1^\dagger \hat{B}_1 + \hbar G_1 (\hat{B}_0 \hat{B}_1^\dagger + \hat{B}_0^\dagger \hat{B}_1) \\ &+ \sum_{i>0} \hbar\tilde{\Omega}_i^{(1)} \hat{\tilde{B}}_i^{(1)\dagger} \hat{\tilde{B}}_i^{(1)} + \hbar\tilde{\lambda}_i^{(1)}\hat{B}_1\hat{\tilde{B}}_i^{(1)\dagger} + \hbar\tilde{\lambda}_i^{(1)*}\hat{B}_1^\dagger \hat{\tilde{B}}_i^{(1)},
    \end{split}
\end{align}
where $G_1 = \sqrt{\sum_{i>0}\abs*{\tilde{\lambda}_i^{(0)}}^2} = \sqrt{\int \dd{\omega}J_\mathrm{res}^{(0)}(\omega)}$ and $\Omega_1 = \int\dd{\omega}\omega J_\mathrm{res}^{(0)}(\omega)/\int\dd{\omega}J_\mathrm{res}^{(0)}(\omega)$.
Thus, from the residual exciton spectral density, we can extract a new collective mode with frequency $\Omega_1$ that couples to $\hat{B}_0$ with strength $G_1$ and to a new residual exciton environment with spectral density generated by Eq.~\eqref{eq:res-spec-dens} (cf. Fig.~\ref{fig:chain-mapping-illustration}b). This process can be iterated indefinitely, thereby generating a one-dimensional chain of coupled modes. We already know that the first residual spectral density {$J_\mathrm{res}^{(0)}(\omega)$} is related to the original exciton spectral density {$J(\omega)$} through Eq.~\eqref{eq:res-spec-dens}. Thus, the iterative procedure of extracting new reaction coordinates generates a sequence of residual spectral densities through the recurrence relation
\begin{align}
J_\mathrm{res}^{(n)}(\omega) =   \frac{G_{n}^2J_\mathrm{res}^{(n-1)}(\omega)}{\Phi_{n-1}^2(\omega) + \pi^2[J_\mathrm{res}^{(n-1)}(\omega)]^2}, \;\;\; n\geq 0{,}
\end{align}
where $J_\mathrm{res}^{(-1)}(\omega):=J(\omega)$, $G_n = \sqrt{\int J_\mathrm{res}^{(n-1)}(\omega)\dd{\omega}}$ is the coupling strength between the modes $n$ and $n-1${,} and
\begin{align}
\Phi_{n}(\omega) =  \frac{1}{2}\lim_{\ell\rightarrow 0^+}\int_a^b J_\mathrm{res}^{(n)}(\nu)\qty[\frac{1}{\omega-\nu-{\text{i}}\ell} + \frac{1}{\omega-\nu+{\text{i}}\ell}].
\end{align}
The corresponding mode frequencies are $\Omega_n = \int J_\mathrm{res}^{(n-1)}(\omega)\omega\dd{\omega}/ \int J_\mathrm{res}^{(n-1)}(\omega)\dd{\omega}$\rev{, and the} 
Hamiltonian at the $n$'th iteration of the prodedure is thus
\begin{align}
  \begin{split}
    \hat{H}_n &=  \hbar{\omega_\text{c}} \hat{a}^\dagger \hat{a} + \hbar G_0 (\hat{a}^\dagger \hat{B}_0 + a \hat{B}_0^\dagger) + \hat{W}\\ &+ \sum_{i=0}^n \hbar\Omega_i \hat{B}_i^\dagger \hat{B}_i + \sum_{i=0}^{n-1} \hbar G_{i+1} (\hat{B}_i^\dagger \hat{B}_{i+1} + \hat{B}_i \hat{B}_{i+1}^\dagger) + H'_n,
    \end{split}
\end{align}
where
\begin{align}
  \hat{H}_n' = \sum_{j} \tilde{\Omega}_j^{(n)}\hat{B}_j^{(n)\dagger}\hat{\tilde{B}}_j^{(n)} +  \tilde{\lambda}_j^{(n)}\hat{B}_n \hat{B}_j^{(n)\dagger}  + \tilde{\lambda}_j^{(n)*}\hat{B}_n^\dagger \hat{\tilde{B}}_j^{(n)}
\end{align}
describes interactions with the $n$'th residual exciton environment.

\begin{figure}
  \centering
  \includegraphics[width=\columnwidth]{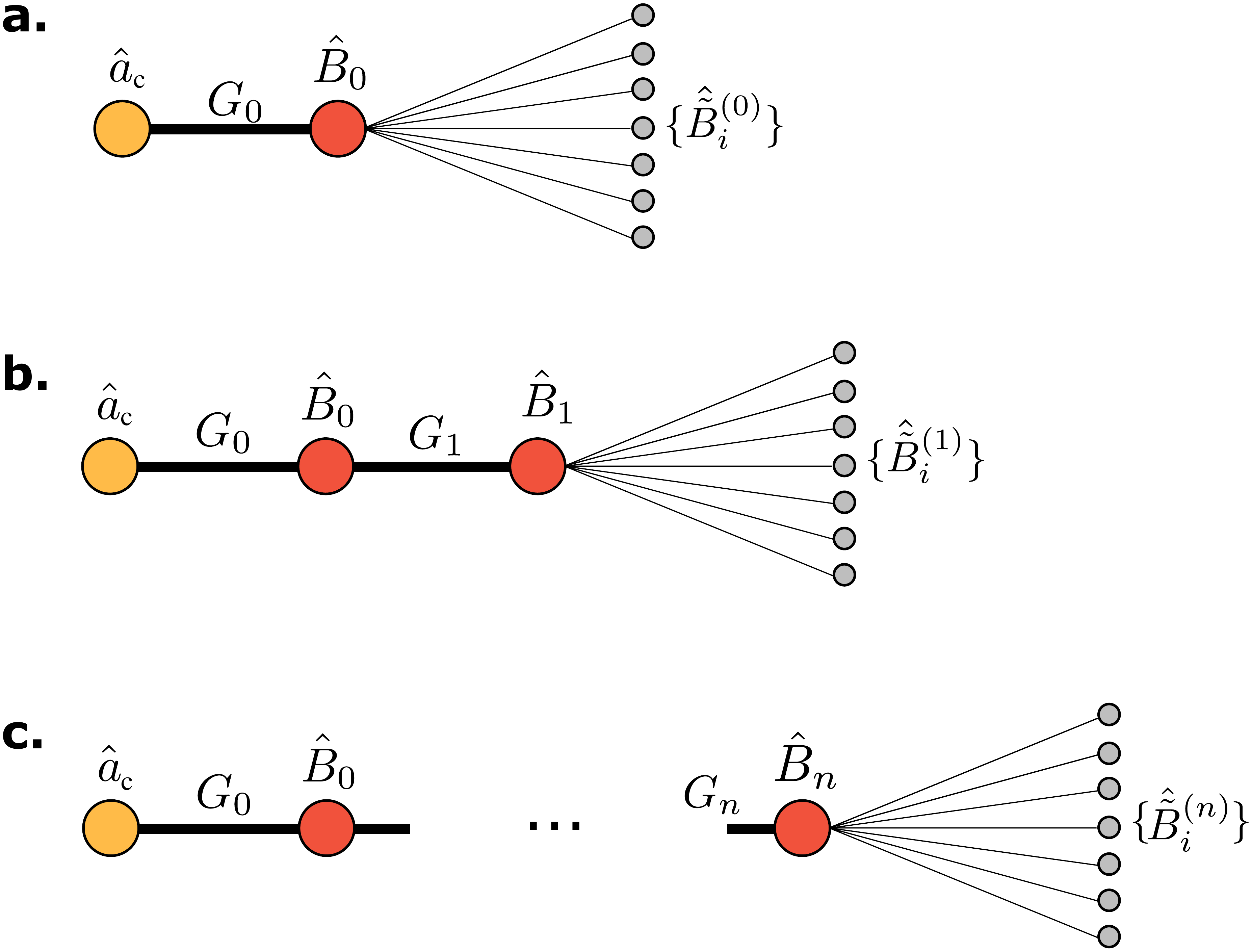}
  \caption{Illustration of the chain mapping technique. {\bf a.} Schematic of the single-mode reaction coordinate mapping, which generates a single reaction coordinate $(\hat{B}_0)$ coupled to a residual exciton environment with mode operators $\hat{\tilde{B}}_i^{(0)}$. {\bf b.} When the reaction coordinate mapping is repeated, a new reaction coordinate, $B_1$, is extracted from the residual environment. This reaction coordinate is then coupled to a new residual environment with operators $\hat{\tilde{B}}_i^{(1)}$. {\bf c.} After $n$ repetitions of the mapping, the excitons are represented as a chain with $n+1$ bosonic modes and a residual excitonic environment.}
  \label{fig:chain-mapping-illustration}
\end{figure}

A {possible} strategy for making the non-Markovian time evolution tractable is to neglect the $n$th residual environment, thereby truncating the chain at the $n$th level,
\begin{align}
\tilde{H}_n = \hat{H}_n - \hat{H}'_n.
\end{align}
This procedure converges towards the exact result as $n\rightarrow\infty${, and} 
the number of chain links required to obtain an error that is low enough for the method to be useful depends on the specific system studied. For example, it was previously found that structures with Fano interference effects are very challenging to capture with a one-dimensional chain representation~\cite{denning2019quantum}. Furthermore, the error of a particular $n$-truncation depends on the time scale over which one is interested in the dynamics: If the chain is initially unpopulated, population will flow from the resonator mode through the chain. This is the relevant situation in the present case because the initial state is the thermal state at or below room temperature where there are no excitons. At longer times, a larger portion of the chain is explored by non-vanishing populations and thus more chain links are necessary to resolve the evolution.
In the time evolution of the truncated chain system {we} also include dissipation {of the electromagnetic field} 
and excitonic line broadening effects of the reaction coordinate as discussed in Sec.~\ref{sec:incl-excit-broad}, by evolving the density operator of the {resonant field} 
and $n$ chain modes with the master equation
\begin{align}
\begin{split}
\dv{\hat{\rho}}{t} = &-\frac{{\text{i}}}{\hbar}[\tilde{H}_n, \hat{\rho}] + 2\gamma_{\text{c}}\mathcal{D}(\hat{a},\hat{\rho})\\ & + 2\gamma_{\text{x}}\mathcal{D}(\hat{B}_0,\hat{\rho}) + 2\gamma_{\text{x}}'\mathcal{D}(\hat{B}_0^\dagger\hat{B}_0,\hat{\rho}){.}
\end{split}
\end{align}

Although {beyond the scope of this article,} 
we note that the one-dimensional chain is amenable to numerically efficient and exact renormalisation-group methods~\cite{prior2010efficient,white1992density,schollwock2005density,vidal2004efficient,daley2004time,white2004real}. Here we shall discuss how the time evolution within a limited time window can be accurately captured by using a truncated chain mapping with a finite number of sites.

\subsubsection{Benchmarking}
\label{sec:chain-benchmarking}

\begin{figure}
  \centering
  \includegraphics[width=\columnwidth]{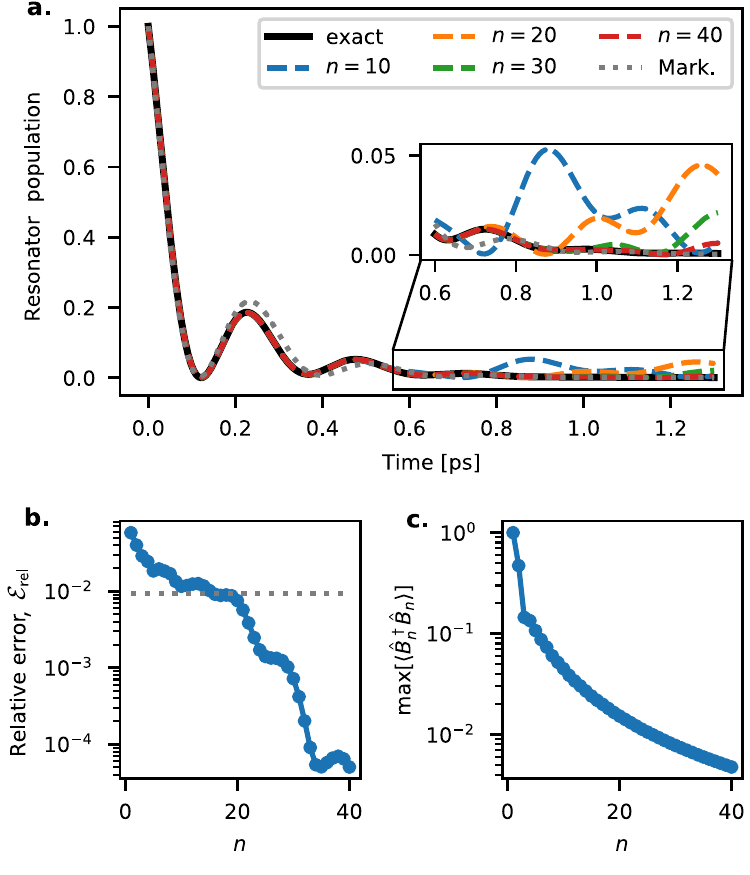}
  \caption{Comparison between truncated chain mapping and exact solution for $\mathrm{WS_2}$ coupled to a Gaussian resonator mode with ${L}=4\mathrm{\; nm}, L_z=200 \mathrm{\; nm}, 2\hbar\gamma_{\text{c}}=6.6\mathrm{\; meV}$. {\bf a.} Time evolution of the resonator population for the exact solution (black solid) and the chain mapping with truncation after $n$ links (dashed lines) for different values of $n$. For reference, the time evolution calculated using the Markovian master equation, Eq.~\eqref{eq:markovian-me-simple} is also shown (grey dotted line). {\bf b.} Relative error of the $n$-truncated chain mappings as a function of $n$. The dotted grey line indicates the error of the Markovian master equation, Eq.~\eqref{eq:markovian-me-simple}. {\bf c.} Maximum population of the last chain site over the evolution time span as a function of the truncation length $n$. }
  \label{fig:chain-vs-exact}
\end{figure}

To test the precision of a given truncation {$n$}, we can consider the situation in Sec.~\ref{sec:exact-evol-single}, where the drive is turned off and the electromagnetic resonator is initialised in a single-excitation state. In this case, we can benchmark the truncated chain expansion against the exact time evolution and evaluate the error. An example of such a comparison is presented in Fig.~\ref{fig:chain-vs-exact}. The time evolution in Fig.~\ref{fig:chain-vs-exact}a demonstrates that a chain with more links allows {one} to evolve the system further in time before the error becomes {pronounced}. 
This is further supported in Fig.~\ref{fig:chain-vs-exact}b, which shows that the relative error of the truncated chain mapping{, as defined in Eq.~(\ref{Eq:rel_error_calc}),} decreases when more links are included. Here, the error of the Markovian master equation, Eq.~\eqref{eq:markovian-me-simple} is indicated with a dotted grey line, demonstrating that it is possible to go below this error and thus resolve non-Markovian effects in the residual exciton environment. Additionally, Fig.~\ref{fig:chain-vs-exact}c shows that the maximum population of the last chain link during the evolution time decreases monotonically with the number of chain links. To carry out this analysis, we have neglected excitonic line broadening effects by setting $\gamma_{\text{x}}$ and $\gamma_{\text{x}}'$ to zero, since these effects are not compatible with the exact strategy described in Sec.~\ref{sec:exact-evol-single}. We do, however, note that the influence of the residual exciton environment is expected to be less important when additional decay channels are present. As such, we should think of the error in Fig.~\ref{fig:chain-vs-exact}b as an upper bound on the truncation error that is expected when excitonic line broadening effects are included.

For reference, the numerical benchmark calculations of the Markovian master equations and the chain-mapped master equation have also been performed with a larger resonator linewidth and at higher temperature of 300 K, which means that the phonon-induced broadening is more pronounced. These calculations can be found in Appendix~\ref{sec:alternative-time-evolution-benchmark} and show the same overall behaviour as the calculations presented here. However, the effect of the residual excitons is seen to be smaller, which is attributed to the increased dissipation into the other decay channels, i.e. through resonator losses and phonon-induced exciton decay.

\subsubsection{Pulsed driving in the linear regime}
\label{sec:pulse-driving-linear}
\begin{figure}
  \centering
  \includegraphics[width=\columnwidth]{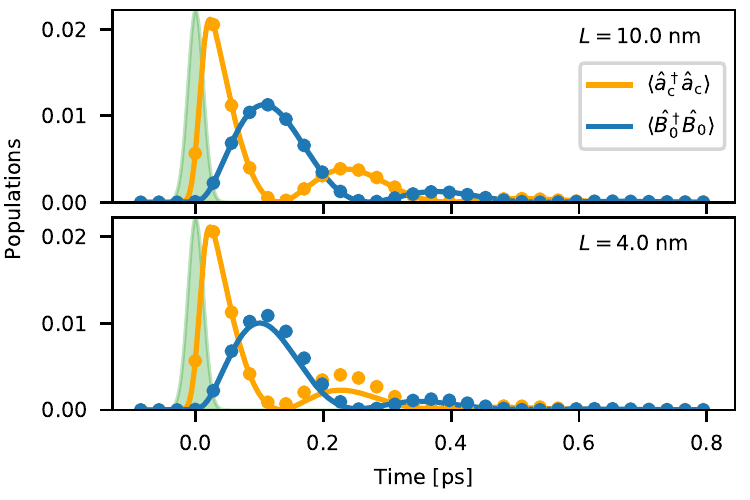}
  \caption{Response to a short driving pulse in the linear regime for $\mathrm{WS_2}$ coupled resonantly $({\omega_\text{c}}=\Omega_0)$ to a Gaussian optical mode with $L=10\mathrm{\; nm}$ and 4 nm, respectively, and $L_z=200 \mathrm{\; nm}, \; 2\hbar\gamma_{\text{c}}=6.6\mathrm{\; meV}, \;n_x^2+n_y^2=0.2$. The temperature is set to 4 K, leading to $\hbar\gamma_{\text{x}}=2.1\mathrm{\; meV},\; \hbar\gamma_{\text{x}}'=0.11\mathrm{\; meV}$, and the optical pulse parameters are $\Delta = 0.2/G_0,\; A = 0.1/\Delta$ and $\; \omega_{\text{d}}=\Omega_0={\omega_\text{c}}$. The resonator (orange) and exciton reaction coordinate (blue) populations for the full system calculated with an $n=30$ chain mapping are shown with solid lines. The corresponding time evolution obtained by ignoring the residual excitons altogether is shown with dots.}
  \label{fig:pulse-drive-linear}
\end{figure}

One of the benefits of the chain-mapping technique is the simplicity of introducing driving with a time-dependent amplitude, $F(t)$. For the Markovian master equation derived in Sec.~\ref{sec:markovian-master-equation}, this would result in a decay term, $\mathcal{K}$, which is explicitly time dependent. Here, we use the chain mapping to study the response of the exciton-resonator system to a short laser pulse that weakly perturbs the system. By making sure that the initially induced population of the resonant field is far below unity, the nonlinear exciton--exciton interaction can be neglected, while the interactions with the residual exciton modes are captured by the chain of bosonic modes. The Hamiltonian for the truncated chain in a frame rotating at the driving frequency is
\begin{align}
  \begin{split}
    &\hat{\tilde{H}}_n(t) =   \hbar F(t)(\hat{a}_{\rm c}+\hat{a}_{\rm c}^\dagger) + \hbar({\omega_\text{c}}-\omega_{\text{d}}) \hat{a}_{\rm c}^\dagger \hat{a}_{\rm c} \\ &+ \hbar G_0 (\hat{a}_{\rm c}^\dagger \hat{B}_0 + \hat{a}_{\rm c} \hat{B}_0^\dagger) \\ &+ \sum_{i=0}^n \hbar(\Omega_i-\omega_{\text{d}}) \hat{B}_i^\dagger \hat{B}_i + \sum_{i=1}^n \hbar G_i (\hat{B}_i^\dagger \hat{B}_{i-1} + \hat{B}_i \hat{B}_{i-1}^\dagger),
    \end{split}
\end{align}
and we shall focus on a Gaussian pulse amplitude of the form 
\begin{align}
F(t) = A e^{-(t-t_0)^2/\Delta^2}.
\end{align}
We also re-instate the excitonic line broadening effects described in Sec.~\ref{sec:incl-excit-broad} with the rates $\gamma_{\text{x}}$ and $\gamma'_{\text{x}}$.

In Fig.~\ref{fig:pulse-drive-linear}, the dynamics of the {electromagnetic field} 
(orange solid) and exciton reaction coordinate (blue solid) is plotted for two different lateral length scales of the electromagnetic resonator and compared to the corresponding evolution obtained by ignoring the residual excitons (dots). As the lateral length scale decreases, the exciton cutoff frequency $\xi$ increases as $1/L^2$, effectively increasing the interaction between the reaction coordinate and the residual exciton modes. For $L=4\mathrm{\;nm}$, we find a relative error of $0.8\%$ in the resonator field evolution and $0.4\%$ for the exciton reaction coordinate evolution, when the residual excitons are ignored. This means that we can justify the use of a Markovian theory with a single exciton mode when $L>4\:\mathrm{nm}$ for the given parameters. In the next section, we shall investigate the nonlinear response in this regime. As we shall see, there is a parameter regime where $L$ is large enough (here above $4\:{\rm nm}$) to neglect the residual exciton modes, yet small enough that the nonlinear interactions become pronounced.


\subsubsection{Pulsed driving in the nonlinear regime}
\label{sec:pulse-driv-nonl}

\begin{figure}
  \centering
  \includegraphics[width=\columnwidth]{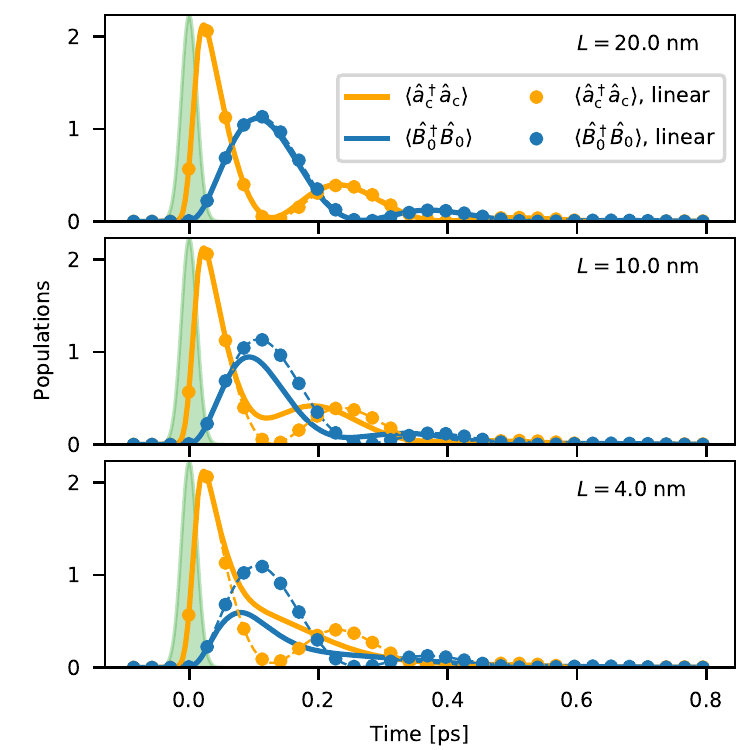}
  \caption{Nonlinear response of the system due to exciton--exciton interactions. The solid lines show the resonator (orange) and exciton reaction coordinate (blue) populations calculated from the full Hamiltonian including exciton--exciton interactions. The dots and dashed lines show the corresponding time evolution when the exciton--exciton interaction is neglected. Parameters: $A=1/\Delta$, and otherwise as in Fig.~\ref{fig:pulse-drive-linear}.}
  \label{fig:pulse-drive-nonlinear}
\end{figure}
{To investigate the 
nonlinear response of the system,} 
we include exciton interactions in the local exciton reaction coordinate through the term $\hat{W}_0$, as described in Sec.~\ref{sec:excit-excit-inter}. To see the nonlinear response, we need to increase the {amplitude $A$ of}  
the driving pulse. 
In practice, this means that we need to resolve a higher number of excitations in the system{, whereby} 
it becomes significantly more challenging to include a higher number of chain sites. {At this point, therefore,} 
we choose the lateral size of the mode to be sufficiently large that the residual modes can be neglected (${L}\geq 4\mathrm{\; nm}$ for the parameters in Fig.~\ref{fig:pulse-drive-linear}). As before, we show in Fig.~\ref{fig:pulse-drive-nonlinear} the time evolution of the system (solid lines) for different lateral optical confinement lengths, $L$. {We} 
compare the evolution with the linear response, {as} obtained by removing the interaction term $\hat{W}_0$ from the Hamiltonian (dots, dashed lines), thereby uncovering the role of exciton--exciton interactions in the dynamical evolution. Due to the increase of the nonlinearity as $L$ is decreased, a larger deviation between the full dynamics and the linear response is observed. The dominating nonlinear effect is a reduced energy transfer from the resonant field to the exciton reaction coordinate, {which arises} 
because the transition from one to two excitons is shifted away from resonance due to the nonlinear interactions. This is a signature of a polariton blockade process, i.e. the inhibition of multi-polariton excitations.
\revv{The dynamics in Fig.~\ref{fig:pulse-drive-nonlinear} shows that there exists a parameter regime, where the lateral confinement length is sufficiently small that the nonlinear interactions influence the dynamics, yet large enough that the residual excitons can be ignored.}
In Ref.~\onlinecite{shortmanuscript}, we study the impact of the nonlinear interactions in terms of polariton blockade in detail and establish the conditions for reaching blockade. Most importantly, we find that polariton blockade is reached when the nonlinear interaction strength, $W_0'$, exceeds the polariton dephasing generated by $\gamma_x'$.

\section{Semiclassical limit}
\label{sec:benchm-with-semicl}
As an alternative to the reaction-coordinate approach to exciton--resonator interactions, we now develop a semiclassical description of the interaction of the electromagnetic field with the excitons in terms of the excitonic dielectric response. This allows us to connect the fundamental material parameters to the dielectric function, which has been experimentally measured for several materials~\cite{li2014measurement,zhang2014absorption}. Furthermore, it allows us to carry out an independent {classical} reference calculation{, which we expect to agree with the result of our microscopic theory in the weak-excitation limit of linear response. This reference calculation thus serves as an important consistency check of the microscopic theory.} 

\subsection{Exciton susceptibility}
\label{sec:diel-resp-excit}

In a purely classical framework, we can model the electromagnetic response of a two-dimensional material in the plane $z=z_0$ as a thin polarizable sheet of thickness $d$ with a relative permittivity distribution given by
\begin{align}
\epsilon_\text{R}(\mathbf{r},\omega) =
\begin{cases}
1+\chi(\omega) \;&\text{for} \; |z-z_0|< d/2,\\
1 \;&\text{otherwise}.
\end{cases}
\end{align}
In cases where the sheet is illuminated by an incoming electromagnetic field, we can calculate the total electric field by use of the Lippmann-Schwinger equation. For the present analysis, we can confine the discussion to the case of normal incidence and consider incoming electric fields of the general form $\mE_\text{in}(\mr,\omega) = E_\text{in}(z,\omega)\mathbf{n}$, where $\mathbf{n}$ is a unit polarization vector, taken to be linear for simplicity. The total field can then be calculated as the solution to the equation
\begin{align}
\label{Eq:Lippmann_Schwinger_classical}
\begin{split}
E_\text{tot}(z,\omega) &= E_\text{in}(z,\omega) \\ &+ k_0^2\int_{z_0-d/2}^{z_0+d/2}\dd{z'} G_\text{B}(z,z',\omega)\chi(\omega)E_\text{tot}(z',\omega),
\end{split}
\end{align}
in which $k_0=\omega/c$ is the ratio of the angular frequency to the speed of light, {and $G_\text{B}(z,z',\omega) =  \text{i}\exp[\text{i}k_0|z-z'|]/(2k_0)$} 
is the one-dimensional electric field Green function of the homogeneous background. 
For sufficiently thin materials, we can assume the integrand to be approximately constant, wherefore we can solve the equation \pk{to find} 
\begin{align}
E_\text{tot}({z_0},\omega) =  \frac{E_\text{in}(z_0,\omega)}{1-k_0^2dG_\text{B}({z_0,z_0},\omega)\chi(\omega)}.
\label{Eq:total_field_classical}
\end{align}

In order to establish a link between the fundamental excitonic properties and the corresponding susceptibility, we consider the total field generated by weakly driving the excitons with a set of normal-incidence 
plane waves of the form
\begin{align}
\mathbf{f}_\mu(\mr) =  \frac{\text{e}^{\mathrm{i}\omega_\mu z/\text{c}}}{\sqrt{S Z}}\mathbf{n},
\label{Eq:1Dmodes}
\end{align}
where $Z$ is the depth of the quantization volume. To calculate the total field, we use the full interaction Hamiltonian with non-rotating wave terms, Eq.~\eqref{eq:H-light-matter-2nd-quant}, and follow Refs.~\onlinecite{wubs2004multiple,wubs2002local,wubs2004spontaneous} to derive a 
Lipmann-Schwinger equation for the vector potential operator $\hat{\mathbf{A}}(z,\omega)=\mathbf{n}\hat{A}(z,\omega)$, expressed in terms of the incoming vector potential operator $\hat{\mathbf{A}}^{(0)}(z,\omega)=\mathbf{n}\hat{A}^{(0)}(z,\omega)$,
of the form
\begin{align}
\label{eq:lippmann-schwinger-A}
\begin{split}
&\hat{A}(z,\omega) = \hat{A}^{(0)}(z,\omega) 
+\int\dd{z'} G(z,z',\omega)V(z',\omega)\hat{A}(z',\omega),
\end{split}
\end{align}
where $V(z,\omega) = V_0(\omega)\delta(z-z_0)$, 
\begin{align}
V_0(\omega) = \sum_\alpha\frac{2Se_0^2\abs{\mathbf{p}_{\rm cv}^\alpha\cdot\mathbf{n}}^2}{\pi\hbar\epsilon_0m_0^2c^2a_{\rm B}^2}\qty(\frac{1}{\omega-\omega_0} - \frac{1}{\omega+\omega_0}),
\end{align}
and we have ignored a possible source term for the vector potential, because we consider a scattering problem, where no excitons are initially excited. The Green's function entering Eq.~\eqref{eq:lippmann-schwinger-A} is related to $G_{\rm B}$ as $G(z,z',\omega)=-G_{\rm B}(z,z',\omega)/S$.
We can solve Eq.~\eqref{eq:lippmann-schwinger-A} to find
\begin{align}
\label{eq:L-S-A-solved}
  \hat{A}(z_0,\omega) = \frac{\hat{A}^{(0)}(z_0,\omega)}{1 + G_{\rm B}(z_0,z_0,\omega)V_0(\omega)/S},
\end{align}
and since the electric field is the derivative of the vector potential, we find 
that the same equation holds for the electric field, under the substitutions $\hat{A} \rightarrow \hat{E}$ and $\hat{A}^{(0)}\rightarrow\hat{E}^{(0)}$.

By comparing Eqs.~\eqref{Eq:total_field_classical} and \eqref{eq:L-S-A-solved}, we can directly see that the two expressions are equivalent, when $k_0^2 d\chi(\omega)=-V_0(\omega)/S$. This requirement yields the expression for the exciton susceptibility,
\begin{align}
\chi(\omega) = \sum_\alpha\frac{2 e_0^2\abs{\mathbf{p}_{cv}^\alpha\cdot\mathbf{n}}^2}{\pi\hbar\epsilon_0m_0^2c^2a_{\rm B}^2d}\frac{1}{\omega^2}\qty(\frac{1}{\omega+\omega_0}-\frac{1}{\omega-\omega_0}). 
\end{align}

To connect to susceptibility measurements in the literature, we note that the exciton line broadening due to \pk{non-}radiative decay and phonon interactions should also be taken into account in the susceptibility. We include these effects by a complex shift of the poles of the susceptibility at $\omega=\pm\omega_0$ into the lower part of the complex plane by the total line broadening $\Gamma_\pk{\text{x}}=\gamma_\pk{\text{x}}+\gamma_\pk{\text{x}}'$. The resulting susceptibility, which serves as a model of what one would measure in linear-response measurements, then takes the form
\begin{align}
  \begin{split}
\chi(\omega) = \sum_\alpha&\frac{2 e_0^2\abs{\mathbf{p}_{cv}^\alpha\cdot\mathbf{n}}^2}{\pi\hbar\epsilon_0m_0^2c^2a_{\rm B}^2d}\frac{1}{\omega^2}\\ &\times\qty(\frac{1}{\omega+\omega_0+\pk{\text{i}}\Gamma_\pk{\text{x}}}-\frac{1}{\omega-\omega_0+\pk{\text{i}}\Gamma_\pk{\text{x}}}),
\end{split}
\end{align}
The poles of interest are both located in the lower part of the complex plane, as they should be due to causality of the electromagnetic response in the time domain, and one of the poles is at negative real frequencies, wherefore it can often be neglected. When more exciton transitions are present, a generalization gives an expression with multiple pole terms. Combining these pairwise and dropping small terms of order $\Gamma^2_\pk{\text{x}}$, 
we find the familiar form
\begin{align}
\chi(\omega) = \sum_m \frac{f_m}{\omega_{0,m}^2-\omega^2-2\text{i}\omega\Gamma_{\mathrm{x},m}},
\label{Eq:suceptibility_standard_form}
\end{align}
where the oscillator strengths, $f_m$, are given by
\begin{align}
f_m = \sum_\alpha\frac{4e_0^2 \abs{\mathbf{p}_{\text{cv},m}^\alpha\cdot\mathbf{n}}^2}{\pi \hbar\epsilon_0m_0^2\omega_{0,m}a_\text{B}^2d}.
\label{Eq:Oscillator_strength_micro}
\end{align}
With the assumption that the field polarisation is in the plane, such that $n_x^2 + n_y^2 = 1$, we can write the oscillator strength as
\begin{align}
f_m = \frac{4 e_0^2 p_\mathrm{\mathrm{cv},m}^2}{\pi m_0^2\epsilon_0\hbar\omega_0 a^2_\pk{\text{B}} d}.
\end{align}
This dielectric parameter $f_m$ has been measured for several 2D materials and provides a very useful means to determine the exciton Bohr radius from experiments~\cite{li2014measurement,zhang2014absorption}. For $\mathrm{WS_2}$, using the parameters in Table~\ref{tab:2D-mat-parameters} and the experimentally obtained value for the 1s exciton $\hbar^2f_{0}=1.9\mathrm{\: eV^2}$ from Ref.~\onlinecite{li2014measurement} yields the exciton Bohr radius $a_\mathrm{B}=1.95\mathrm{\; nm}$.

\subsection{Reference calculation}
\label{sec:refer-calc}

\begin{figure}
  \centering
\includegraphics[width=\columnwidth]{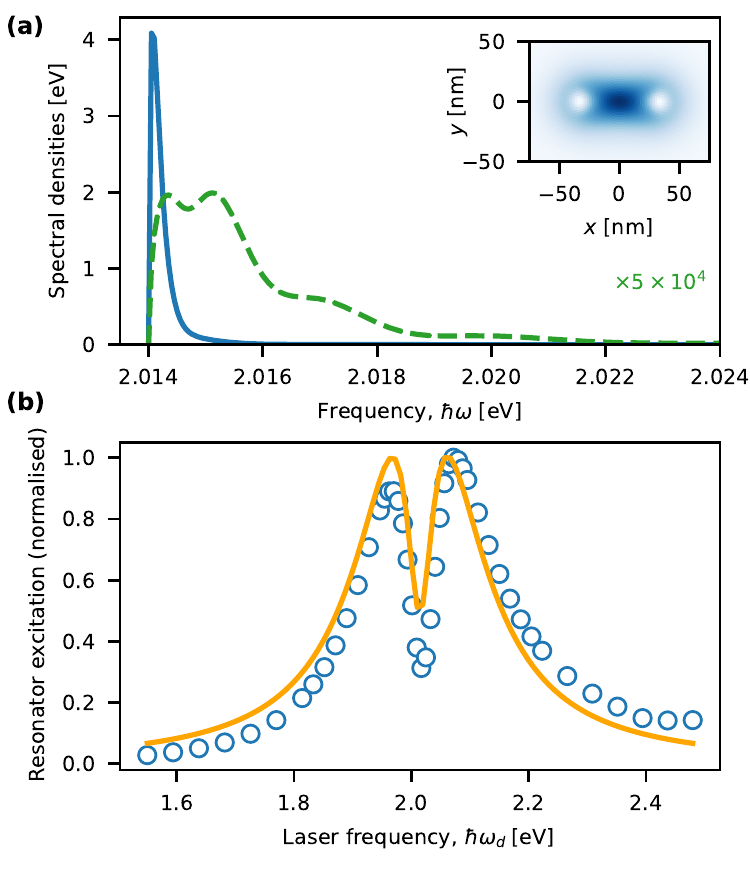}
  \caption{{\bf a.} Exciton spectral density (blue solid) and residual spectral density (green dashed, scaled up by a factor of $5\times 10^4$) for a single quasi-normal mode of a gold nanorod coupled to a monolayer sheet of $\mathrm{WS_2}$, as illustrated in Fig.~\ref{fig:intro}. The projected field profile, $\abs*{{\tilde{F}^x_\text{c}}(\mathbf{r},0)}^2 + \abs*{{\tilde{F}^y_\text{c}}(\mathbf{r},0)}^2$ is shown in the inset. {\bf b.} Excitation spectrum of the system when driven by an external laser field, calculated with semiclassical theory, $I(\omega_\pk{\text{d}})$ (blue circles), and the microscopic quantum model using the Markovian master equation, $n_{\rm ss}(\omega_\pk{\text{d}})$ (orange solid line). In order to compare the semiclassical spectrum (calculated as the  field intensity in the middle of the nanorod) with the quantum spectra (calculated as the steady-state expectation value of the resonator population), the spectra have been scaled with their maximum values. The light-matter coupling strength calculated from the quantum theory (Eq.~\eqref{Eq:G0_real_space_mode_function_integral}) is $\hbar G_0 = \text{35.1 meV}$. }
  \label{fig:nanorod}
\end{figure}

We are now in a position to compare \pk{the microscopic theory} 
to a semiclassical reference calculation. To \pk{this end,} 
we consider a gold nanorod coupled to a monolayer of $\mathrm{WS_2}$ 
and calculate the linear excitation spectrum when driving with an external laser \pk{at different frequencies}. 
\pk{
%
The nanorod is modeled as a cylinder with spherical end caps, as depicted in Fig.~\ref{fig:intro}, a diameter of 30 nm, and a total length of 90~nm. For convenience, we define a coordinate system in which the nanorod is oriented in the $x$-direction, and its center is at the position $(x,y,z)=(0,0,20\mathrm{\:nm})$. For these calculations, we use a Drude permittivity model of the form
\begin{align}
\epsilon_\text{R}(\mr,\omega) = 1 - \frac{\omega_\text{p}^2}{\omega(\omega+\text{i}\gamma)},
\end{align}
with $\hbar\omega_\text{p}=6.9\:\mathrm{eV}$ and $\hbar\gamma=0.2\mathrm{\: eV}$.

The numerical QNM calculations were carried out with the boundary-element method ``MNPBEM''~\cite{Hohenester_JPC_183_370_2012} and with the iterative search method of Ref.~\onlinecite{Alpeggiani_SR_6_34772_2016}, see Ref.~\onlinecite{Kristensen_AOP_12_612_2020} for details. Mesh generation using triangular surface elements was done by the open source mesh generator ``Gmsh''~\cite{gmsh}. The dipolar QNM of interest has a complex resonance frequency of $\tlo_\text{c} \ell_0/2\pi\text{c} = 0.1625(2) - 0.00920(2)\text{i}$. When the field is scaled to unity at the position $\mr_0=(0,0,0)$ the complex inverse norm of the QNM is found to be $\ell_0^3\mft_\text{c}^2(\mr_0)/\langle\langle\mft_\text{c}(\mr)|\mft_\text{c}(\mr)\rangle\rangle=1.246(2)-0.0307(5)\text{i}$. Here, $\ell_0=100\mathrm{\:nm}$ is a fixed length scale that is used to express the QNM frequency and norm in dimensionless units. In the single-QNM approximation, it can be shown that the normalization factor $S_\text{c}$ appearing in the derivations in Ref.~\onlinecite{franke2019quantization} always takes the value $S_\text{c}=1$, which is the value that we used for the calculations in this work. \rev{The} exciton spectral density, $J(\omega)$ and the residual spectral density, $J_\mathrm{res}(\omega)$ (green dashed), are shown in Fig.~\ref{fig:nanorod}a along with the absolute value of the in-plane components of the field profile $\tilde{\mathbf{F}}_{\rm c}$, as derived from the fundamental dipolar QNM and discussed in Appendix~\ref{App:Electric_field_outside_the_resonator}. The dipolar nature of the field profile results in a non-trivial in-plane distribution of the electric field as shown in the inset of Fig.~\ref{fig:nanorod} for the case of $z=z_0$. It follows from the analysis in Sec.~\ref{sec:single-electr-mode} that this is the field distribution defining the exciton reaction coordinate and the coupling constant $G_0$, recall Eq.~(\ref{Eq:G0_real_space_mode_function_integral}).} %
Running a benchmark calculation as described in Sec.~\ref{sec:markov-benchmarking}, we find that the relative error obtained by neglecting the residual exciton modes is \pk{less than one in a thousand, wherefore we neglect the residual excitons in the analysis.} 
We then calculate the excitation spectrum as the steady-state photon number, $n_{\rm ss}(\omega_\pk{\text{d}}):=\Tr[\hat{a}_{\rm c}^\dagger\hat{a}_{\rm c}\hat{\rho}_{\rm ss}(\omega_\pk{\text{d}})]$ where $\hat{\rho}_{\rm ss}(\omega_\pk{\text{d}})$ is the steady-state density operator of Eq.~\eqref{eq:markovian-me-simple} with constant driving (frequency $\omega_\pk{\text{d}}$ and driving strength $F$). \pk{For these calculations, the} driving strength $\hbar F$ \pk{was} set to a very low value of $36\mu\text{eV}$ to ensure that the system is in the linear-response regime.

As a semiclassical reference calculation, we can also calculate the excitation spectrum through a solution of the classical electromagnetic problem, where the $\mathrm{WS_2}$ monolayer is modelled as a sheet of thickness \pk{$d=0.618$nm} with a dielectric function corresponding to Eq.~\eqref{Eq:suceptibility_standard_form} with experimentally measured parameters~\cite{li2014measurement}, $\hbar^2 f_{0} = 1.9\mathrm{\; eV^2}, \; \hbar\omega_0 =2.014\mathrm{\; eV}$, and the exciton linewidth $\hbar\Gamma_{\rm x}=16.1\mathrm{\:meV}$, calculated from Table~\ref{tab:exciton-broadening-parameters} using $T=300\mathrm{\; K}$. \rev{The} reference calculations were done with MNPBEM and the same mesh for the nanorod as was used for the QNM calculations; additional scattering from the 2D material was included by the use of the appropriate background Green function~\cite{Waxenegger_CPC_193_138_2015} corresponding to a thin sheet with a single Drude-Lorentz pole. In practical experiments, the sample will rest on a substrate and there will be additional contributions to the optical response of the $\mathrm{WS_2}$-material and other similar corrections, which will lead to shifts in the resonance frequencies of the nanorod and possibly the 2D material. In Appendix~\ref{sec:substrate}, we investigate the effect of a substrate on the electromagnetic response and show that the nanorod can always be tuned into resonance with the excitons by various means, for example by varying the length of the nanorods. In order to simplify the model and focus on the dominating physics, we have left out these effects here.

The driving laser field was modeled as an incoming plane wave with frequency $\omega_\pk{\text{d}}$, and for each value of this frequency, the resonator excitation \pk{was} measured as the field intensity in the middle of the nanorod, $I(\omega_\pk{\text{d}})$. In order to compare the field intensity $I(\omega_\pk{\text{d}})$ and the steady-state photon number $n_{\rm ss}(\omega_\pk{\text{d}})$, we normalise both to their maximum values and plot them together in Fig.~\ref{fig:nanorod}b. The asymmetry in the semiclassical spectrum can be attributed to a frequency-dependent incoupling factor between the external driving field and the resonator \pk{field} 
term, which can be derived from coupled-mode theory~\cite{Kristensen_AOP_12_612_2020}, but which is not accounted for in the \ptk{present approach.} 
The remaining discrepancy is attributed to the \rev{non-retarded coupling and the} approximation that only a single \pk{QNM} is taken into account in the \pk{microscopic} 
model. \pk{Importantly, we find that the calculated splitting of the spectrum in the two independent calculation methods differ only by 0.5\%\ptk{. In combination with the general qualitative agreement between the two spectra, we interpret this as a demonstration of consistency between the} 
microscopic quantum model \ptk{and the} 
semiclassical theory \ptk{based on} 
measurements of the linear exciton susceptibility. \rev{In this limit of linear response, it is an interesting fact that one can also treat the problem from a purely electromagnetic point of view and model the response by use of two quasi-normal modes, as was recently presented in Ref.~\onlinecite{carlson2021strong}.}

\rev{In closing, we emphasize} that the \rev{general} microscopic 
model is \pk{applicable also} 
beyond the linear, semiclassical regime, when nonlinear effects and few-exciton statistics become important, as \rev{discussed} in Secs.~\ref{sec:excit-excit-inter} and~\ref{sec:pulse-driv-nonl}\rev{, as well as in} 
Ref.~\onlinecite{shortmanuscript}.

\section{Conclusion}
\label{sec:conclusion}

In conclusion, we have developed a microscopic quantum theory for the interaction between \pk{an electromagnetic resonator and} excitons in a \pk{pristine sheet of 2D semiconductor material.} 
In particular, by invoking a basis change of the exciton continuum, we have identified a collective exciton mode, termed the exciton reaction coordinate, that effectively accounts for the light-matter interaction.
We have derived analytic expressions for the coupling strength between the resonant electromagnetic field and the reaction coordinate, thereby showing that it is independent of the lateral confinement of the field. 

To calculate the dynamical evolution of the system, we have introduced and analyzed several Markovian and non-Markovian approaches and assessed their regimes of validity. Using these strategies, we have evaluated the importance of the residual exciton environment, which is coupled to the reaction coordinate. We find that \pk{the influence of the residual excitons} 
becomes more pronounced when the lateral optical mode dimensions become smaller\pk{. In many cases, however, the residual excitons can be ignored altogether.} 
For the extreme regime where the electromagnetic field is laterally confined to a characteristic length scale of a few nanometers, it becomes necessary to account for the residual excitons. We have developed an iterative chain-representation of the residual exciton environment, which is able to resolve non-Markovian effects and thus to go beyond the Markovian master equation.

We have also derived the linear dielectric response of the excitons, which allows \pk{one} to connect the material parameters to the dielectric function and to consistently interface \pk{the microscopic} theory with a semiclassical approach. Furthermore, we have calculated the nonlinear interaction strength of the excitons within the reaction coordinate and found that it scales as the inverse \pk{area of the electromagnetic field in the 2D material}, meaning that laterally 
\pk{confined electromagnetic fields} 
lead to stronger exciton-exciton interactions.
In this context, we have found that there exists an interesting parameter regime, where the lateral confinement length scale is large enough that the residual excitons can be ignored, but small enough that nonlinear effects are significant.

\begin{acknowledgments}
The authors thank \rev{Peder Meisner Lyngby} for valuable discussions and Tony Heinz for providing the experimental data from Ref.~\onlinecite{li2014measurement}. This work was supported by the Danish National Research Foundation through NanoPhoton - Center for Nanophotonics, grant number DNRF147 and Center for Nanostructured Graphene, grant number DNRF103. NS acknowledges support from the Villum Foundation through grant number 00028233. EVD acknowledges support from Independent Research Fund Denmark through an International Postdoc fellowship (grant no. 0164-00014B). M.W. and N.S. acknowledge support from the Independent Research Fund
Denmark - Natural Sciences (project no. 0135-00403B)
\end{acknowledgments}

\appendix

\section{Electric-field operators outside the resonator}
\label{App:Electric_field_outside_the_resonator}
\pk{
Expansions based on QNMs can often provide a good approximation to the electromagnetic field at positions inside or close to electromagnetic resonators. In the present case of a single QNM approximation, in particular, we can expand the electric-field operator as
\begin{align}
\mEh(\mr,\omega) = \text{i}\sqrt{\frac{\hbar\omega_\text{c}}{2\epsilon_0}}\mft_\text{c}(\mr,\tlo_\text{c})\hat{a}_\text{c}(\omega) + \text{H.c.},
\label{Eq:mEh_franke_single_mode}
\end{align}
where $\hat{a}_\text{c}^\dagger$ and $\hat{a}_\text{c}$ are bosonic raising and lowering operators obeying the commutation relation $[\hat{a}_\text{c}(t),\hat{a}_\text{c}^\dagger(t)]=1$~\cite{franke2019quantization}. At positions far away from the resonator, the QNM expansions in general are expected to fail~\cite{Kristensen_AOP_12_612_2020}, and this poses a challenge for the application at hand, which involves infinitely extended sheets of 2D materials. Since the electric-field operator obeys Maxwell's equations, however, we can calculate the field operator at general positions $\mr$ by use of the three-dimensional electric-field equivalent of Eq.~(\ref{eq:lippmann-schwinger-A}),
\begin{align}
\mathbf{\hat{E}}_\text{tot}(\mr,\omega) &= \mathbf{\hat{E}}_0(\mr,\omega)\nonumber\\ &+ \left(\frac{\omega}{\text{c}}\right)^2\int\dd[3]{\mathbf{r'}}\mathbf{G}(\mr,\mr',\omega)\Delta\epsilon(\mr',\omega)\mathbf{\hat{E}}_\text{tot}(\mr',\omega),
\label{Eq:Lippmann_Schwinger_Ehat}
\end{align}
where $\mG(\mr,\mr',\omega)$ is the electric-field Green tensor of the homogeneous background material of permittivity $\epsilon_\text{B}$, and $\Delta\epsilon(\mr,\omega)=\epsilon_\text{R}(\mr,\omega)-\epsilon_\text{B}$ is the change in the relative permittivity defining the electromagnetic resonator.

The first term in Eq.~(\ref{Eq:Lippmann_Schwinger_Ehat}) represents the free-space electric-field operator in the absence of the resonator and therefore does not contribute to the resonant field dynamics that we aim to describe. For these calculations, therefore, we drop this term and rewrite the expression by substituting the QNM expansion of the electric-field operator in Eq.~(\ref{Eq:mEh_franke_single_mode}) as
\begin{align}
\mEh(\mr,\omega) = \text{i}\sqrt{\frac{\hbar\omega_\text{c}}{2\epsilon_0}}\mFt_\text{c}(\mr,\omega)\hat{a}_\text{c}(\omega),
\label{Eq:Ehat_omega_retarded}
\end{align}
where
\begin{align}
\mFt_\text{c}(\mr,\omega) = \left(\frac{\omega}{\text{c}}\right)^2\int_V\dd[2]{\mathbf{\mr'}}\mG(\mr,\mr',\omega)\Delta\epsilon(\mr',\omega)\mft_\text{c}(\mr')
\label{Eq:BigF}
\end{align}
is the analytical continuation of the electric field QNM onto the real axis~\cite{Ge_NJP_16_113048_2014}. Equation (\ref{Eq:Ehat_omega_retarded}) represents the fully retarded electric-field operator pertaining to the field of interest in the electromagnetic resonator. In the temporal dynamics, the retardation becomes explicitly evident as the convolution in Eq.~(\ref{Eq:mEh_time_retarded_convolution}). When coupling to very localized excitons, however, we can simplify the expression considerably by evaluating $\mFt_\text{c}(\omega)$ at $\omega = \omega_\text{c}$ to focus on the instantaneous response only. In the same spirit, we restrict the analysis to the local dynamics by replacing $\mG(\mr,\mr',\omega)$ in Eq.~(\ref{Eq:BigF}) by the quasistatic Green tensor. In doing so, we ensure that the integral defining the coupling strength in Eq.~(\ref{Eq:G0_real_space_mode_function_integral}) is convergent.

}


\section{Exciton dissipator}
\label{sec:exciton-dissipator}
The derivation of the exciton dissipator follows the standard approach as described in detail in Ref.~\onlinecite{breuer2002theory}. In this Appendix, we apply the approach to the present situation, where the resonator and exciton reaction coordinate are treated as an open quantum system, which is coupled to an environment consisting of the residual exciton modes.
  Starting from Eq.~(\ref{Eq:exciton_dissipator_starting_point}), 
we decompose the interaction Hamiltonian as {$\hat{H}_{\text{SR}} = \hbar(\hat{S}_1\hat{R}_1 + \hat{S}_2\hat{R}_2)$}, where 
\begin{align}
\begin{split}
  \hat{S}_1&=\hat{B}_0^\dagger, \;\;\; \hat{R}_1 =\sum_{i>0}\tilde{\lambda}_i^*\hat{\tilde{B}}_i 
\end{split}
\end{align}
and $\hat{S}_2=\hat{S}_1^\dagger,\; \hat{R}_2=\hat{R}_1^\dagger$.
To \pk{simplify the expression for the} 
interaction-picture time evolution\pk{,} 
we introduce the eigenstate-projected system operators
\begin{align}
  \label{eq:80}
  \hat{S}_i(\omega)&:=\sum_{\mathcal{E}'-\mathcal{E}=\omega}\Pi(\mathcal{E})\hat{S}_i\Pi(\mathcal{E}'),
\end{align}
\pk{with $i\in\{1,2\}$,} where $\Pi(\mathcal{E})$ is the projector onto the system subspace with eigenenergy $\mathcal{E}$ with respect to $\hat{H}_\pk{\text{S}}$,
\begin{align}
  \label{eq:82}
  \Pi(\mathcal{E})&=\sum_{\omega_l=\mathcal{E}} \dyad{l}, \; \hat{H}_\pk{\text{S}}\ket{l}=\hbar \omega_l\ket{l}.
\end{align}
We 
\rev{distinguish} between $\hat{S}_i^\dagger(\omega)$ and $[\hat{S}_i(\omega)]^\dagger$, such that the former expression refers to $\sum _{\mathcal{E}'-\mathcal{E}=\omega} \Pi(\mathcal{E})\hat{S}_i^\dagger\Pi(\mathcal{E}')$ and the latter to $\sum_{\mathcal{E}'-\mathcal{E}=\omega} [\Pi(\mathcal{E})\hat{S}_i\Pi(\mathcal{E}')]^\dagger=\hat{S}_i^\dagger(-\omega)$. The interaction-picture time evolution of the projected system operators is then
\begin{align}
  \label{eq:83}
  e^{\pk{\text{i}}\hat{H}_\pk{\text{S}}t/\hbar}\hat{S}_i(\omega)e^{-\pk{\text{i}}\hat{H}_\pk{\text{S}}t/\hbar}&=\hat{S}_i(\omega)e^{-\pk{\text{i}}\omega t}\pk{.}
\end{align}
\pk{Using the} 
completeness of the eigenstates of $\hat{H}_\pk{\text{S}}$, we find
 \begin{align}
   \label{eq:84}
   \hat{S}_i&=\sum_\omega \hat{S}_i(\omega)=\sum_\omega \hat{S}_i(-\omega), \\ \hat{S}_i(t) &= \sum_\omega \hat{S}_i(\omega)e^{-\pk{\text{i}}\omega t}=\sum_\omega \hat{S}_i(-\omega)e^{\pk{\text{i}}\omega t}\pk{,}
 \end{align}
\pk{and the} 
dissipator due to the residual excitons can thus be written as
\begin{align}
  \label{eq:85}
\begin{split}
  \mathcal{K}[\hat{\rho}]&=-\sum_{ij}\sum_{\omega\omega'} \int_0^\infty\dd{\tau}\Lambda_{ij}(\tau)e^{\pk{\text{i}}\omega'\tau}[\hat{S}_i(-\omega),\hat{S}_j(\omega')\hat{\rho}]\\ & + \int_0^\infty\dd{\tau}\Lambda_{ji}(-\tau)e^{-\pk{\text{i}}\omega'\tau}[\hat{\rho} \hat{S}_j(-\omega'),\hat{S}_i(\omega)],
\end{split}
\end{align}
where $\Lambda_{ij}(\tau) = \Tr_\pk{\text{R}}\qty{\hat{R}_i e^{-\pk{\text{i}}\hat{H}_R\tau/\hbar}\hat{R}_je^{+\pk{\text{i}}\hat{H}_R\tau/\hbar}\hat{\rho}^0_\pk{\text{R}}}$ is a residual excitonic correlation function. Here, $\hat{\rho}_{\rm R}^0$ is the initial density operator of the residual exciton environment, which is taken to be the vacuum state as a good approximation to the thermal state of a semiconductor.
As described in Ref.~\onlinecite{breuer2002theory}, a so-called secular approximation is enforced by \pk{keeping} only 
terms with $\omega=\omega'$ in the summation\pk{, which is justified by the fact that a factor of $\exp\{\pk{\text{i}}(\omega-\omega')t\}$ appears in the sum for} 
the interaction-picture time evolution of the reduced density operator\pk{; if} 
$\omega\neq\omega'$, the exponential is assumed to average out to zero. \pk{In addition to simplifying the expression,} 
the secular approximation \pk{ensures} 
that the dynamics generated by the master equation is completely positive and trace preserving~\cite{breuer2002theory,de2017dynamics}. Noting that, due to $\hat{\rho}_{\rm R}^0$ being the vacuum state, the only nonzero correlation function is $\Lambda_{12}(\tau)=\sum_{i>0}|\tilde{\lambda}_i|^2\pk{\exp}\{-\pk{\text{i}}(\tilde{\Omega}_{i}-\omega_\pk{\text{d}})\tau\}$, the secularised residual exciton dissipator becomes
\begin{align}
\label{eq:secular-markovian-dissipator-appendix}
\begin{split}
  \mathcal{K}[\hat{\rho}] = -\sum_\omega &\Big\{\Gamma_\mathrm{res}(\omega)\mathcal{D}[\hat{B}_0(\omega),\hat{\rho}]\\ & - \pk{\text{i}}\Delta_\mathrm{res}(\omega)[ (\hat{B}_0(\omega))^\dagger \hat{B}_0(\omega),\hat{\rho}]\Big\},
\end{split}
\end{align}
where
\begin{align}
  \Gamma_\mathrm{res}(\omega) &= 2\Re\qty{\int_0^\infty\dd{\tau} \Lambda_{12}(\tau)e^{i\omega\tau}}, \\ \Delta_\mathrm{res} (\omega) &= \Im\qty{\int_0^\infty\dd{\tau} \Lambda_{12}(\tau)e^{i\omega\tau}}.
\end{align}
The \pk{second term in Eq.~(\ref{eq:secular-markovian-dissipator-appendix})} 
amounts to a shift of the resonance energies and will be neglected here. The remaining part describes exciton dissipation with a rate that can be written in terms of the residual spectral density as in Eq.~(\ref{eq:secular-markovian-decay-rate}).
}

\section{Effect of dielectric substrate}
\label{sec:substrate}
\begin{figure}
\centering
\includegraphics[width=\columnwidth]{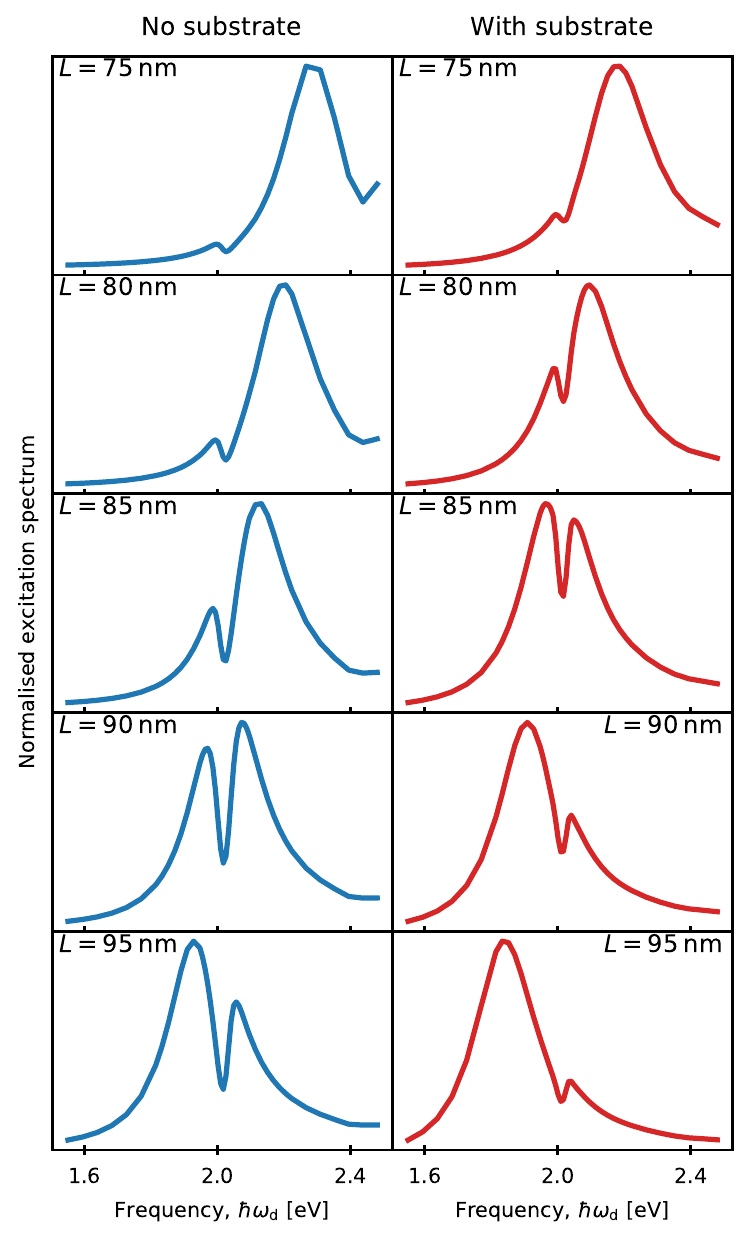}
  \caption{Semiclassical excitation spectrum of gold nanorod coupled to monolayer $\mathrm{WS_2}$ as in Fig.~\ref{fig:nanorod}, calculated without (left panel, blue) and with (right panel, red) dielectric substrate, and for nanorod lengths between 75 nm and 95 nm as indicated with text.}
  \label{fig:substrate}
\end{figure}

In order to simplify the model and highlight the do\-minating physics, the reference calculations in Section~\ref{sec:refer-calc} were performed for a gold nanorod above a thin sheet of material characterized by a single Drude-Lorentz pole. Additional corrections to the model will serve primarily to shift the resonance frequency of the nanorod or the excitonic transitions. These effects, therefore, are not so different from unknown perturbations in practical experiments, which can be compensated by tuning of the material system to bring it into resonance. In Fig.~\ref{fig:substrate}, we illustrate how such a tuning can be performed by changing the nanorod length, similar to the approach of Wen et al.~\cite{wen2017room,geisler2019single}. The left panel of Fig.~\ref{fig:substrate} shows calculations identical to those in Fig.~\ref{fig:nanorod} of the main text, except for the use of nanorods of different lengths \rev{ranging from $L=75\mathrm{\; nm}$ to $L=95\mathrm{\; nm}$}. Clearly, by changing the length of the nanorods, one is able to tune the system into resonance. The right panel of Fig.~\ref{fig:substrate} shows the situation when the system is changed by introducing a substrate with permittivity $\epsilon_\text{subs} = 2.12$ extending infinitely downwards from just below the thin sheet of two-dimensional material. Notably, we did not include an encapsulation layer, since typical experiments of this sort are performed without~\cite{wen2017room,zheng2017manipulating,kleemann2017strong, cuadra2018observation, stuhrenberg2018strong, han2018rabi, geisler2019single, qin2020revealing}. Furthermore, the full experimentally measured response of $\mathrm{WS_2}$ is included with all poles, corresponding to not only the lowest-lying A1s-exciton, but also the higher-lying exciton states, as detailed in Ref.~\onlinecite{li2014measurement}. In this case, the resonance condition has changed, so that it is now fulfilled by nanorods of approximately 85 nm length, but the general anti-crossing trend in the curves is unchanged, since the dominating physics is still that of two strongly coupled harmonic oscillators.

\section{Variation of coupling strength for nanorods of different length}
\label{sec:variation-nanorod-length}

\begin{figure}
\centering
\includegraphics[width=\columnwidth]{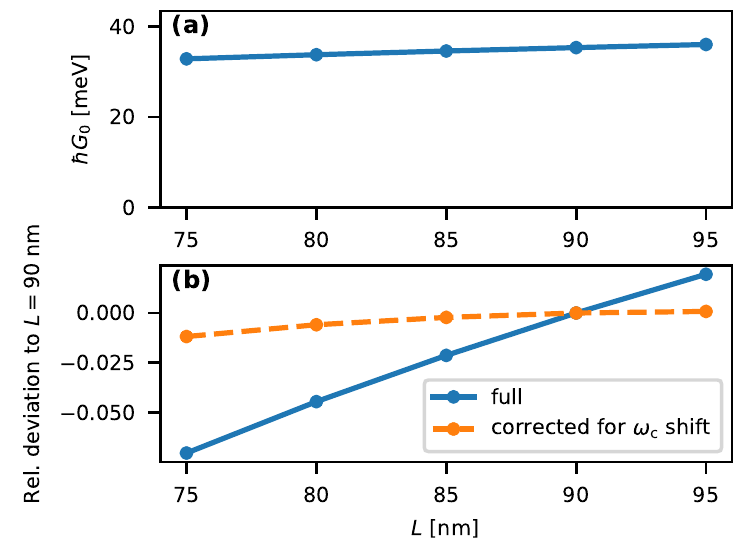}
  \caption{{\bf a.} Coupling strength $G_0$ of a nanorod resonator coupled to monolayer $\mathrm{WS_2}$ as in Sec.~\ref{sec:refer-calc}, for varying lengths of the nanorod, $L$. {\bf b.} Relative deviation of the coupling strength as compared to $L=90\mathrm{\; nm}$. The blue data points and lines show the deviation of the raw coupling strength, $G_0$, whereas the orange datapoints and lines show the coupling strength corrected for the shift in resonance frequency that accompanies the change in the resonator length, $\sqrt{\omega_{\rm c}}G_0$.}
  \label{fig:nanorod-size-variation}
\end{figure}

\rev{In order to substantiate the 
claim that the coupling strength $G_0$ is largely independent of the lateral confinement length scale, we explicitly compare the value of $G_0$ for gold nanorods of different lengths ranging from $L=75\mathrm{\; nm}$ to $L=95\mathrm{\; nm}$, but otherwise identical to the one that was} investigated in Sec.~\ref{sec:refer-calc}\rev{, see also App.~\ref{sec:substrate}}. Fig.~\ref{fig:nanorod-size-variation}a shows $G_0$ as a function of the nanorod length, $L$, and the relative difference compared to $L=90\mathrm{\;nm}$ is shown in Fig.~\ref{fig:nanorod-size-variation}b (blue data points and lines). While the coupling strength \rev{increases as a function of length,} 
this effect is mainly due to the change in the resonance frequency $\omega_{\rm c}$\rev{, which decreases with increasing length, as seen in Fig.~\ref{fig:substrate}}. 
In Eq.~\eqref{Eq:G0_real_space_mode_function_integral}, a factor of $1/\sqrt{\omega_{\rm c}}$ appears in the coupling strength. Thus, to make a meaningful comparison of the effect of the spatial mode distribution on the coupling strength, we should multiply the coupling strength by $\sqrt{\omega_{\rm c}}$ to correct for the shift in resonance frequency. This comparison is shown in Fig.~\ref{fig:nanorod-size-variation}b with orange data points and lines and reveals that the change in resonator length by 20\% generates a vanishing shift in $\sqrt{\omega_{\rm c}} G_0$ of around 1\%.

\section{Comparison of time-evolution methods with increased dissipation and decoherence}
\label{sec:alternative-time-evolution-benchmark}

\begin{figure}
\centering
\includegraphics[width=\columnwidth]{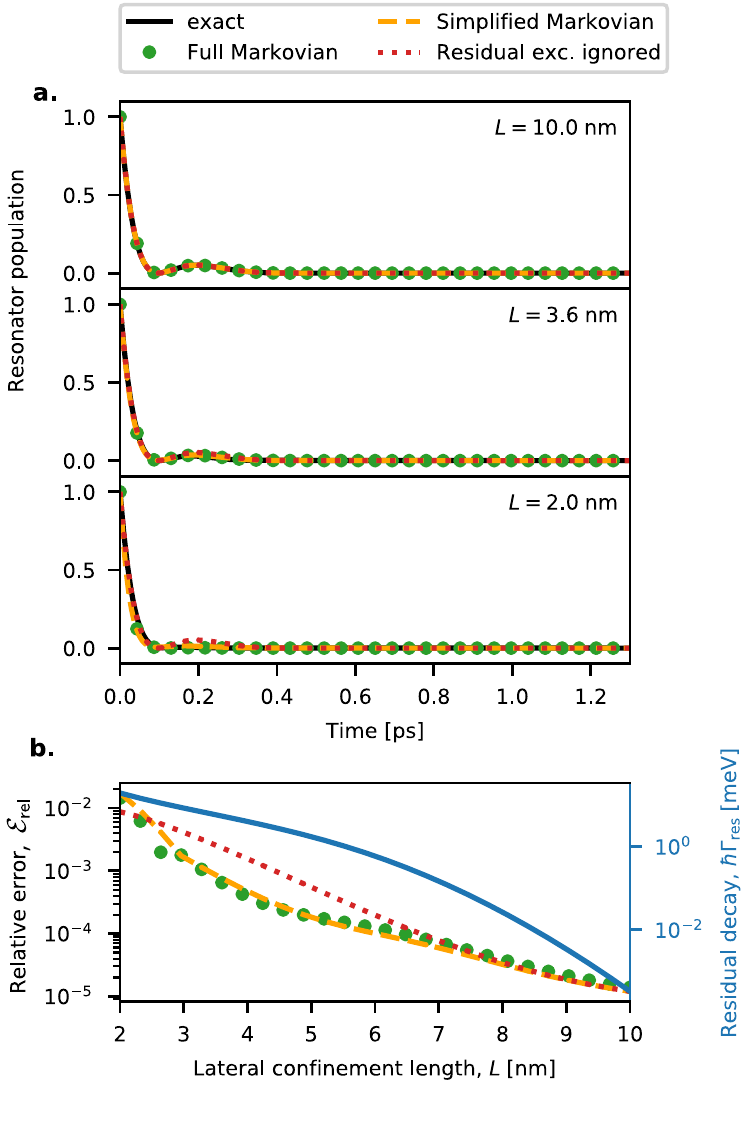}
  \caption{Error calculation of the Markovian master equations corresponding to Fig.~\ref{fig:Markovian-benchmark}, but with the resonator decay rate increased to $2\hbar\gamma_{\rm c}=20\mathrm{\;meV}$.}
  \label{fig:Markovian-benchmark-alternative-parameters}
\end{figure}

\begin{figure}
\centering
\includegraphics[width=\columnwidth]{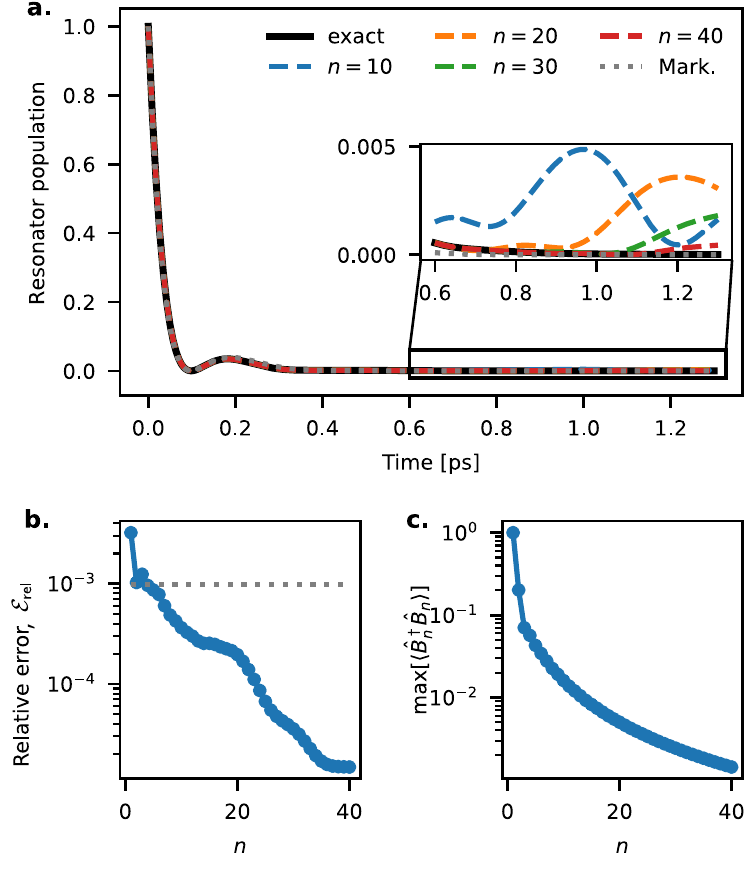}
  \caption{Error calculation of the chain-mapped master equation corresponding to Fig.~\ref{fig:chain-vs-exact}, but with the resonator decay rate increased to $2\hbar\gamma_{\rm c}=20\mathrm{\;meV}$.}
  \label{fig:chain-vs-exact-alternative-parameters}
\end{figure}

\begin{figure}[h]
\centering
\includegraphics[width=\columnwidth]{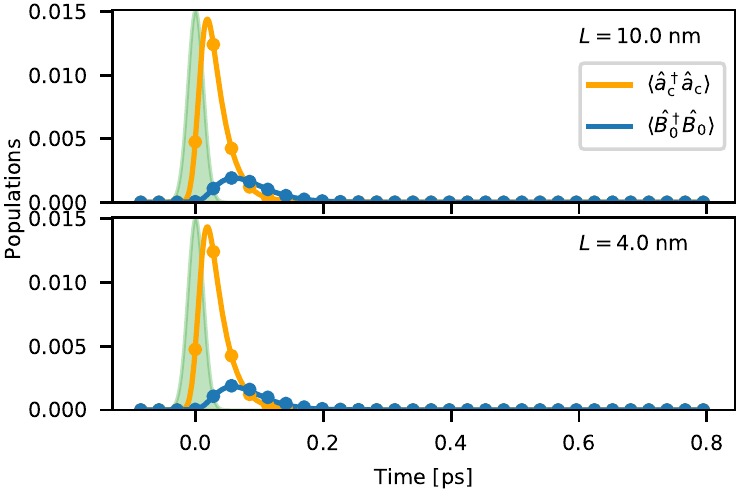}
  \caption{Time evolution calculated with the chain-mapped master equation for pulsed driving, compared to the case where the residual excitons are ignored, as in Fig.~\ref{fig:pulse-drive-linear}, but with the resonator decay rate increased to $2\hbar\gamma_{\rm c}=20\mathrm{\;meV}$ and the temperature increased to 300 K, leading to the phonon-induced exciton decay rate $\hbar\gamma_{\rm x}=7.7\mathrm{\; meV}$ and decoherence rate $\hbar\gamma_{\rm x}^{\:\prime}=8.4\mathrm{\;meV}$.}
  \label{fig:pulse-drive-linear-alternative-parameters}
\end{figure}

In Sec.~\ref{sec:time-evolution}, three different approaches for calculating the time evolution of the exciton-resonator system were presented and compared, in order to assess their validity. For completeness, we present the same comparison calculations \rev{with the only difference that} 
the resonator decay rate has been increased to $2\hbar\gamma_{\rm c}=20\mathrm{\;meV}$, and the temperature has been increased to 300 K\rev{; the resulting} 
phonon-induced exciton decay rate is $\hbar\gamma_{\rm x}=7.7\mathrm{\; meV}$, and the dephasing is $\hbar\gamma_{\rm x}^{\:\prime}=8.4\mathrm{\;meV}$. \rev{Figure~\ref{fig:Markovian-benchmark-alternative-parameters}} 
corresponds to Fig.~\ref{fig:Markovian-benchmark} and shows the comparison of the Markovian master equations with the exact calculation. \rev{Figure~\ref{fig:chain-vs-exact-alternative-parameters}} 
corresponds to Fig.~\ref{fig:chain-vs-exact} and shows the comparison of the chain-mapped master equation with the exact calculation. We remind that excitonic line broadening due to phonon interactions is not included in these benchmark calculations, since the exact memory-kernel equation, Eq.~\eqref{eq:exact-time-evolution}, is incompatible with these effects. \rev{Figure~\ref{fig:pulse-drive-linear-alternative-parameters}} 
corresponds to Fig.~\ref{fig:pulse-drive-linear} and shows the comparison between the chain-mapped master equation and the case where the residual exciton environment is neglected for pulsed driving. Here, phonon-induced broadening has been included.

\end{document}